\documentclass[12pt]{article}

\usepackage{hyperref}
\usepackage{cite}
\usepackage{units}
\usepackage{epsfig}
\usepackage{pifont}
\usepackage{amsmath,amsfonts,amssymb,amsthm,mathrsfs}
\usepackage{graphicx,chngcntr,slashed}
\usepackage{cancel}
\usepackage{pstricks}
\usepackage{afterpage}
\usepackage{braket}

\def\mathswitch#1{\relax\ifmmode#1\else$#1$\fi}
\newcommand{\msbar}{\mathswitch{\overline{\text{MS}}}}

\newcommand{\mycaption}[1]{\caption{\sl #1}}

\makeatletter
\def\section{\@startsection {section}{1}{\z@}{+3.0ex plus +1ex minus
  +.2ex}{2.3ex plus .2ex}{\large\bf\boldmath}}
\def\subsection{\@startsection{subsection}{2}{\z@}{+2.5ex plus +1ex
minus +.2ex}{1.5ex plus .2ex}{\normalsize\bf\boldmath}}
\def\subsubsection{\@startsection{subsubsection}{3}{\z@}{+3.25ex plus
 +1ex minus +.2ex}{1.5ex plus .2ex}{\normalsize\it}}

\graphicspath{{.}{plots/}}

\begin{document}
\thispagestyle{empty}

\def\thefootnote{\fnsymbol{footnote}}

\begin{flushright}
\end{flushright}

\vspace{1cm}

\begin{center}

{\Large {\bf Top quark decay at next-to-leading order in the Standard Model
  Effective Field Theory}}
\\[3.5em]
{\large
Radja~Boughezal$^1$, Chien-Yi~Chen$^2$, Frank~Petriello$^{1,2}$ and Daniel~Wiegand$^{1,2}$
}

\vspace*{1cm}

{\sl
$^1$ HEP Division, Argonne National Laboratory, Argonne, Illinois 60439, USA \\[1ex]
$^2$ Department of Physics \& Astronomy, Northwestern University,\\ Evanston, Illinois 60208, USA
}

\end{center}

\vspace*{2.5cm}

\begin{abstract}

We consider top quark decay in the Standard Model Effective Field
Theory (SMEFT). We present a calculation of the total decay width and
the $W$-boson helicity fractions at next-to-leading order (NLO) in
SMEFT.  Our result includes the complete set of contributing
four-fermion operators in addition to QCD dipole operators and
bottom-mass suppressed effects.  We show that operators that first
appear at NLO in the SMEFT can
be bounded by the current data as well as future data from both a
high-luminosity LHC and a potential $e^+e^-$ collider, demonstrating the importance of going
beyond leading order when studying the SMEFT.  We discuss technical
aspects of our calculation that we believe will be useful in future
higher-order studies of the SMEFT, in particular the treatment of
$\gamma_5$ in loop diagrams.

\end{abstract}

\setcounter{page}{0}
\setcounter{footnote}{0}

\newpage


\section{Introduction}

\noindent
The Standard Model (SM) has so far been remarkably successful in
describing all data coming from the LHC. In some channels,
percent-level comparisons between theory and data are now possible.
Understanding how indirect signatures of new physics are constrained
by these exquisite data is becoming of central importance in
extracting the most possible information from the LHC, especially
given the lack of new states beyond the SM so far.  This will become
part of the legacy of the LHC program, similar to how the global
electroweak fit became a legacy of the LEP collider.  The appropriate
theoretical framework for investigating these constraints is the SM
effective field theory (SMEFT) containing higher-dimensional operators
formed from SM fields.  The leading dimension-6 operators
characterizing deviations from the SM have been
classified~\cite{Buchmuller:1985jz, Grzadkowski:2010es} (there is a
dimension-5 operator that violates lepton number which we do not
consider).  There has been considerable effort in performing global
analyses of the available data within the framework of SMEFT~\cite{Han:2004az,Pomarol:2013zra,Chen:2013kfa,Ellis:2014dva,Wells:2014pga,Falkowski:2014tna,deBlas:2016ojx}.

Given the precision of the available data, it is critical to address
whether the theoretical predictions entering SMEFT analysis are
sufficiently precise.  There are two primary considerations to
address: whether higher-order corrections containing dimension-6
operators in the SMEFT are necessary, and whether dimension-8 operators
should be considered.  We will consider the first of these issues in
this manuscript.  Within the Standard Model, theoretical corrections
to next-to-next-to-leading order (NNLO) in the QCD coupling constant
and next-to-leading order (NLO) in the electroweak coupling constant
are known for a host of interesting processes.  The situation is 
less advanced in the SMEFT.  NLO results assuming a subset of
contributing SMEFT operators are known for a host of Higgs
decays~\cite{Hartmann:2015oia,Hartmann:2015aia,Dedes:2018seb,Gauld:2015lmb,Gauld:2016kuu,Dawson:2018pyl,Dawson:2018liq}
and $Z$-boson decays~\cite{Hartmann:2016pil,Dawson:2018jlg,Dawson:2018dxp}, as well
as certain Higgs production processes~\cite{Degrande:2012gr,Vryonidou:2018eyv}.  Especially given
the data precision, going beyond leading order is necessary to
properly understand bounds on the SMEFT operators, as has been argued
in the literature (see, for example, Ref.~\cite{Berthier:2015oma,Passarino:2016pzb}).

In this manuscript we study NLO corrections in the SMEFT to top quark
decay.  We focus on the total width and $W$-boson helicity fraction
observables.  There have been several analyses of constraints on the SMEFT
arising from top quark data~\cite{Greiner:2011tt,Zhang:2012cd,Buckley:2015lku,Cirigliano:2016nyn,AguilarSaavedra:2018nen,Hartland:2019bjb}.
NLO QCD corrections to top quark decay in the SMEFT, augmented by the
one-loop contribution from the top-quark chromomagnetic operator, have
been considered~\cite{Zhang:2014rja}.  The precision of the top-quark
data coming from the LHC warrants these detailed investigations of
top-quark properties in the SMEFT.  Our goals in this manuscript are
summarized below.
\begin{itemize}

  \item We extend the previous calculations of higher-order corrections to
    top-quark decays in the SMEFT to also include the bottom-quark
      chromomagnetic dipole operator and all contributing four-Fermi
      operators, both four-quark and semi-leptonic types.  This is a
      further step toward a complete next-to-leading order calculation
      of top-quark decay within the
      SMEFT, which we believe will eventually be warranted by the
      high-luminosity LHC program.  It is also of phenomenological
      interest to determine whether third generation four-quark
      operators can be constrained by this measurement.  Previously
      suggested probes of these operators have focused on production
      of four external heavy-flavor states~\cite{Azzi:2019yne}.

    \item We emphasize the role of chiral Ward identities as an
      important calculational check, in particular in the treatment of
      $\gamma_5$ in loops containing four-fermion operators.  We
      consider several different schemes for the treatment of
      $\gamma_5$ in dimensional regularization and demonstrate how
      imposing chiral Ward identities renders them consistent.  We
      believe that this discussion will be useful in the future as
      higher-order effects in the SMEFT are further studied.
      
    \item We study the effect of bottom-quark mass-suppressed
      contributions in the SMEFT.  Interestingly, such effects go like $m_b/m_t$ in
      the SMEFT at LO due to the chiral structure of the contributing dimension-6
      operators, unlike in the SM where they go as $(m_b/m_t)^2$.
      This leads to significant constraints on these operators from
      current data.

    \item We consider the constraints on all operators using the
      currently available data on the total width and $W$-helicity
      fractions.  We also derive simple projections for a
      high-luminosity LHC and
      a possible future $e^+e^-$ machine.  Our primary goal is to determine how well
      loop-induced operators can be probed given both current and future
      experimental measurements.  As we consider only a subset of the available data rather than
      perform a global fit as
      in~\cite{Buckley:2015lku,AguilarSaavedra:2018nen,Hartland:2019bjb},
      our numerical results should only be considered
      representative of the potential bounds.
      
\end{itemize}

Our paper is organized as follows.  In Section~\ref{sec:SMEFT} we
provide a brief overview of the SMEFT operators relevant for our
calculation of top-quark decay.  We present the leading-order
calculation of our observables in Section~\ref{sec:LO}, establishing
our calculational framework.  We present
our NLO calculation in Section~\ref{sec:NLO}.  We discuss in detail
the technical aspects of the calculation particular to the SMEFT such
as the treatment of $\gamma_5$ and the ultraviolet renormalization.
Our numerical results are shown in Section~\ref{sec:numerics}.
Finally, we conclude in Section~\ref{sec:conc}.


\section{Overview of top-quark decay in SMEFT}
\label{sec:SMEFT}

\begin{figure}[htbp]
\centering
\includegraphics[width=2.0in]{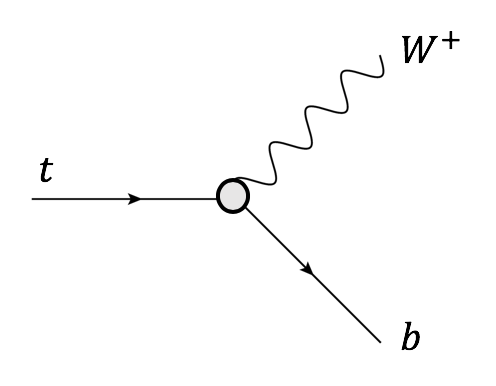}
\vspace{-2ex}
\mycaption{Leading-order Feynman diagram for the process $t\rightarrow Wb$. Through the shaded vertex the Wilson coefficients $C_{tW}, C_{bW}$ and $C_{tb\phi}$ enter the amplitude.
\label{fig:BornFeynman}}
\end{figure}

We begin by discussing the features of the dimension-6 SMEFT relevant to our
calculation of top-quark decay, $t \to Wb$. At leading-order this
process proceeds through the single Feynman diagram shown in
Fig.~\ref{fig:BornFeynman}.
To determine the SMEFT contributions to this
process we use the Warsaw basis~\cite{Buchmuller:1985jz}. Following
the notation of Ref.~\cite{Dedes:2017zog} we find the following
operators contributing at leading-order:
\begin{equation}
\begin{split}
{\cal O}_{\substack{uW \\ pr}}& = \bar{q}_p \sigma^{\mu\nu} u_r
\tau^I\tilde{\phi}W^I_{\mu\nu}, \\
{\cal O}_{\substack{dW \\ pr}}& = \bar{q}_p \sigma^{\mu\nu} d_r
\tau^I\phi W^I_{\mu\nu}, \\
{\cal O}_{\substack{\phi ud \\ pr}} &= i (\tilde{\phi}^{\dagger} D_{\mu} \phi )
(\bar{u}_p \gamma^{\mu}d_r).
\end{split}
\end{equation}
Here, $q_p$ denotes the left-handed quark doublet with $p$ the
generation index, $u_r$ and $d_r$ are respectively the up and down
right-handed singlet quarks with generation index $r$, $\phi$ is the
Higgs doublet, and $W^I_{\mu\nu}$ is the field-strength tensor for the
$SU(2)_L$ gauge bosons with $I$ denoting the isospin index.
$\sigma_{\mu\nu}$ is written in terms of the commutator of $\gamma$ matrices as
  $\sigma_{\mu\nu} = i[\gamma_{\mu},\gamma_{\nu}]/2$. These
operators are written in the flavor eigenstate basis. Rotating to the
mass basis introduces mixing matrices into the Wilson coefficient
matrices multiplying these operators. In our analysis we restrict
ourselves to third-generation couplings, and study the operators
\begin{equation}
{\cal O}_{tW} = {\cal O}_{\substack{uW \\ 33}}, \;\; {\cal O}_{bW} =
{\cal O}_{\substack{dW \\ 33}} , \;\; {\cal O}_{\phi tb} = {\cal O}_{\substack{\phi ud \\ 33}} .
\end{equation}
We label the Wilson coefficients multiplying these operators as
$C_{tW}$, $C_{bW}$ and $C_{\phi tb}$ respectively and assume for
simplicity that they are real.  We factor out the energy scale
$1/\Lambda^2$ associated with thes operators being dimension-6 so that
the Wilson coefficients $C_i$ are dimensionless.

In order to illustrate the sensitivity of top-quark decay properties
to NLO effects in SMEFT we consider a subset of the operators that
contribute at NLO. As we will show later explicitly, and as can be seen using
the renormalization group equations of SMEFT, a consistent
NLO calculation using the operators above requires the following QCD
dipole operators:
\begin{align}\label{eq:dipole}
{\cal O}_{\substack{ug \\ pr}}& = \bar{q}_p \sigma^{\mu\nu} T^A u_r \tilde{\phi}G^A_{\mu\nu}, \\
{\cal O}_{\substack{dg \\ pr}}& = \bar{q}_p \sigma^{\mu\nu} T^A d_r \phi G^A_{\mu\nu}.
\end{align}
Here, $G^A_{\mu\nu}$ is the gluon field-strength tensor and $T^A$ are
the color matrices in the fundamental representation. We again
restrict our analysis to third-generation couplings and study the
operators
\begin{equation}
{\cal O}_{tg} = {\cal O}_{\substack{ug \\ 33}}, \;\; {\cal O}_{bg} =
{\cal O}_{\substack{dg \\ 33}}
\end{equation}
We also consider the following four-Fermi operators which potentially
contribute to top decay as well.  In the case of a four-quark operators we have
\begin{align}\label{eq:4quark}
&{\cal O}_{qq}^{(1)} = (\bar{q}_p \gamma^\mu q_r)(\bar{q}_s \gamma_\mu q_t) \;\;\;\;\;\;\;\;
&&{\cal O}_{qq}^{(3)} = (\bar{q}_p \gamma^\mu\tau^a q_r)(\bar{q}_s \gamma_\mu \tau^a q_t)\nonumber\\
&{\cal O}_{ud}^{(1)} = (\bar{u}_p \gamma^\mu u_r)(\bar{d}_s \gamma_\mu d_t)
&&{\cal O}_{ud}^{(8)} = (\bar{u}_p \gamma^\mu T^A u_r)(\bar{d}_s \gamma_\mu T^A d_t)\nonumber\\
&{\cal O}_{qu}^{(1)} = (\bar{q}_p \gamma^\mu q_r)(\bar{u}_s \gamma_\mu u_t)
&&{\cal O}_{qu}^{(8)} = (\bar{q}_p \gamma^\mu T^A q_r)(\bar{u}_s \gamma_\mu T^A u_t)\nonumber\\
&{\cal O}_{qd}^{(1)} = (\bar{q}_p \gamma^\mu q_r)(\bar{d}_s \gamma_\mu d_t)
&&{\cal O}_{qd}^{(8)} = (\bar{q}_p \gamma^\mu T^A q_r)(\bar{d}_s \gamma_\mu T^A d_t)\nonumber\\
&{\cal O}_{quqd}^{(1)} = (\bar{q}_p^j u_r)\epsilon_{jk}(\bar{q}_s^k d_t)
&&{\cal O}_{quqd}^{(8)} = (\bar{q}_p^j T^A u_r)\epsilon_{jk}(\bar{q}_s^k T^A d_t),
\end{align}
where $\tau^a$ are the Pauli matrices and summation over the $SU(2)$
index $a$ is implied. For simplicity we consider only flavor-diagonal
operators in our study; constraints on flavor-violating operators are
generally better obtained from other observables than those considered
here.  We also include the following semi-leptonic
four-fermion operators in our analysis:
\begin{align}\label{eq:semilep}
{\cal O}_{lq}^{(3)} = (\bar{l}_p \gamma^\mu\tau^a l_r)(\bar{q}_s \gamma_\mu \tau^a q_t) \;\;\;\;\;\;\;\;
{\cal O}_{lequ}^{(3)} = (\bar{l}_p^j \sigma^{\mu\nu} e_r)\epsilon_{jk}(\bar{q}^k_s \sigma_{\mu\nu} u_t).
\end{align}
It is necessary for the one-loop renormalization of SMEFT to include
the following operator as well:
\begin{align}
{\cal O}_{\phi q}^{(3)} = i\big(\Phi^\dagger \overleftrightarrow{D}^a_\mu \Phi\big)\big(\bar{q}_p\tau^a \gamma^\mu q_r\big),
\end{align}
which leads to a redefinition of the $CKM$ matrix 
\begin{align}
K_{CKM}\rightarrow K_{CKM}(1+\frac{v^2}{\Lambda^2}C_{\phi q}^{(3)}).
\end{align}
At leading order this generates a term proportional to the SM Born-level matrix element.\\
As pointed out
in~\cite{Jenkins:2013zja,Jenkins:2013wua,Alonso:2013hga}, the poles
associated with the corrections from four-Fermi operators to the
$tbW$-vertex are removed by renormalizing $C_{\phi q}^{(3)}$.  This leads us to the following Lagrangian describing top decay:
\begin{equation}
  {\cal L} = {\cal L}_{SM}+ \frac{1}{\Lambda^2} \sum_i C_i {\cal O}_i
  \label{eq:Lagrangian}
\end{equation}
with $i$ running over all operators previously discussed. All Feynman rules arising
from these operators can be found in Ref.~\cite{Dedes:2017zog}.


\section{Leading-order calculation}
\label{sec:LO}

We discuss here some basic features of our calculation and present
results for the leading-order (LO) top decay width and helicity fractions
in the SMEFT. The only diagram mediating the decay at tree-level is shown
in Fig.~\ref{fig:BornFeynman}.  It is straightforward to derive the amplitude for the decay
$t(p_t) \to b(p_b)W(p_W)$ using the operators of Section~\ref{sec:SMEFT}. We
consider four observables: the total top decay width $\Gamma_\textrm{tot}$, and the decay
fractions into longitudinal, positive and negative $W$-boson helicities. To
obtain these quantities from the decay amplitude it is convenient to
replace the $W$-boson polarization vectors in the squared amplitude
according to
\begin{equation}
\sum \epsilon^{\mu}(p_W) \epsilon^{*\nu}(p_W) = P^{\mu\nu}.
\end{equation}
We use the following projection operators~\cite{Fischer:2000kx}:
\begin{align}
\label{eq:projectors}
P_{\textrm{tot}}^{\mu\nu} & =
-g^{\mu\nu}+\frac{p_W^{\mu}p_W^{\nu}}{M_W^2},\nonumber \\ 
P_{\textrm{L}}^{\mu\nu} &= \frac{ [M_W^2 p_t^{\mu}-p_t\cdot p_W
  p_W^{\mu}][M_W^2 p_t^{\nu}-p_t\cdot p_W p_W^{\nu}]}{M_W^2 m_t^2
  |\vec{p}_W|^2},\nonumber \\ 
P_\textrm{F}^{\mu\nu} &= -\frac{i}{m_t |\vec{p}_W|}
\epsilon^{\mu\nu\sigma\rho}p_{t\sigma}p_{W\rho},\nonumber \\ 
P_{\pm}^{\mu\nu} &=\frac{1}{2} \left\{
  P_{\textrm{tot}}^{\mu\nu}-P_{\textrm{L}}^{\mu\nu} \pm P_\textrm{F}^{\mu\nu} \right\}.
\end{align}
$\vec{p}_W$ denotes the three-momentum of the $W$-boson. We use these
projectors in both the LO and NLO calculations.

In our calculation we only include terms linear in the EFT couplings,
as terms proportional to EFT couplings squared are of the same order
as neglected dimension-8 operators. We include finite bottom-mass
effects at leading order. We present below the top decay width and
helicity fractions in the SMEFT.  For simplicity of presentation we have
expanded them to linear order in $m_b/m_t$ (in our numerical
analysis we keep the LO results to all orders in this ratio):
\begin{align}
\Gamma_\textrm{tot} &= \frac{\overline{g}(x_W^2-1)m_t}{64\pi x_W^{2}}\Big[(x_W^2-1) \big(\overline{g}(1+ 2x_W^2) + 12 \sqrt{2} C_{tW} m_t^2 x_v x_W^2\big)\nonumber\\
&+6 m_t^2 x_b x_v x_W^2 \big(C_{\phi tb} \overline{g} x_v + 2 \sqrt{2} C_{bW} (1 + x_W^2)\big)\Big] + \mathcal{O}(x_b^2),\nonumber\\
F^\textrm{LO}_{\textrm{L}} &= \frac{\Gamma_{\textrm{L}}}{\Gamma_\textrm{tot}} = \frac{8 \sqrt{2} m_t^2 x_v \left(x_W^2-1\right) x_W^2
    C_{tW}+\bar{g} \left(2 x_W^2+1\right)}{\bar{g}
  \left(2 x_W^2+1\right)^2} \nonumber \\ &+\frac{4 \sqrt{2}
   m_t^2 x_v x_b x_W^2 \left(2 C_{bW}
   \left(x_W^2+1\right)+ \frac{\bar{g} x_v}{\sqrt{2}} C_{\phi tb}
   \left(x_W^2+1\right)\right)}{\bar{g}
   \left(2 x_W^2+1\right)^2} +{\cal O}(x_b^2),\nonumber \\ 
F^\textrm{LO}_- &= \frac{\Gamma_{-}}{\Gamma_\textrm{tot}} = \frac{2x_W^2\left(\bar{g} + 2 \bar{g} x_W^2 + 4 \sqrt{2} C_{tW} m_t^2 x_v (1 - x_W^2)\right)}{\bar{g}(1+2x_W^2)^2} \nonumber\\
&+\frac{2 m_t^2 x_b x_v x_W^2 (4 \sqrt{2} C_{bW} x_W^2 (2 + x_W^2) - C_{\phi tb} \bar{g}  x_v (1 - 4 x_W^2))}{\bar{g} (1 + 3 x_W^2 - 4 x_W^6)} +{\cal O}(x_b^2).
\end{align}
We have abbreviated $x_i = m_i/m_t$ and $x_v = v/m_t$, where $v$ is
the Higgs vacuum expectation value (vev). $\Gamma_i$ denotes the partial decay widths for the different W polarizations. $\bar{g}$ is the scaled electroweak coupling required
to canonically normalize the gauge fields in
SMEFT~\cite{Dedes:2017zog}. We note that the positive helicity
fraction can be obtained using the relation $F_+ = 1-F_- - F_{L}$.
This relation can be easily seen to hold using the projectors of
Eq.~(\ref{eq:projectors}). These quantities reduce to the known SM
results~\cite{Fischer:2000kx} when all SMEFT Wilson coefficients are set to zero.

One interesting feature of these results is their dependence on the
bottom-quark mass fraction $x_b$. In the SM the dependence on the
bottom quark mass at Born-level begins at ${\cal O}(x_b^2)$, while in SMEFT it
begins at  ${\cal O}(x_b)$. This is because the EFT operators ${\cal
  O}_{bW}$ and ${\cal O}_{\phi tb}$ have a $V+A$ helicity structure
instead of the $V-A$ structure of the SM. We will see the effect of
this parametric difference in our numerical results.


\section{Next-to-leading-order calculation}
\label{sec:NLO}

\begin{figure}[h!]
\centering
\includegraphics[width=6.5in]{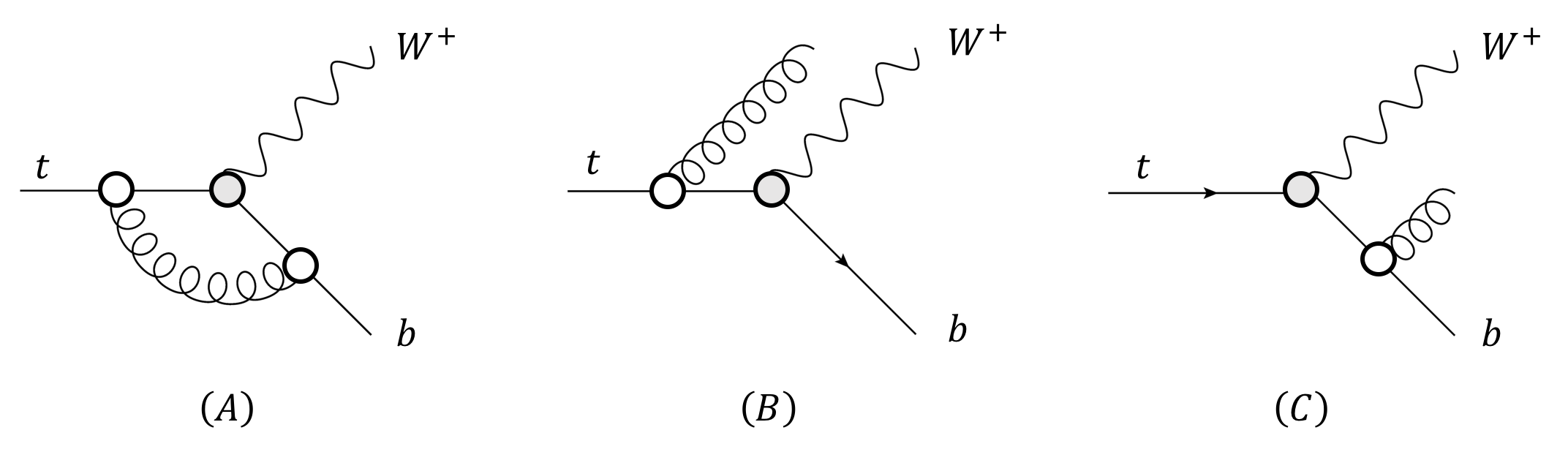}
\vspace{-2ex}
\mycaption{Feynman diagrams for the SMEFT QCD corrections to $t\rightarrow Wb$. Through the white vertices the Wilson coefficients $C_{tg}$ and $C_{bg}$ enter the amplitude. The expansion in $\frac{1}{\Lambda}$ happens at the squared matrix element level.
\label{fig:QCDFeynman}}
\end{figure}

\begin{figure}[h!]
\centering
\includegraphics[width=5.5in]{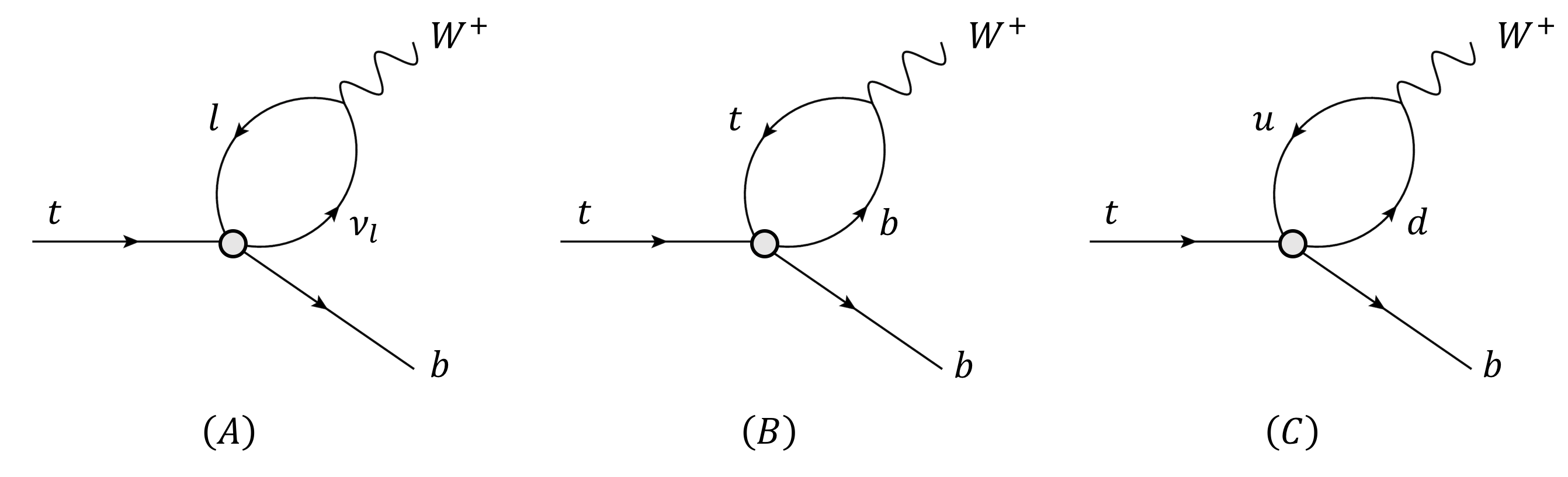}
\vspace{-2ex}
\mycaption{Feynman diagrams for the four-Fermi SMEFT corrections to
  $t\rightarrow Wb$. Through the shaded vertices the Wilson
  coefficients enter the amplitude. The vertex involving the $W$-boson
  receives in principle contributions from SMEFT operators as well,
  but since the four-Fermi vertex has a vanishing SM limit those would lead to terms of order $\frac{1}{\Lambda^4}$ and are therefore neglected.
\label{fig:4FermiFeynman}}
\end{figure}

We discuss in this section our calculation of the NLO corrections to
top-quark decay properties in the SMEFT. Higher-order QCD-like
corrections involving gluon exchange are mediated by both SM QCD and the operators listed in
Section~\ref{sec:SMEFT} giving rise to the diagrams in Fig.~\ref{fig:QCDFeynman}.
Contributions from the four-Fermi operators listed in
Eqs.~(\ref{eq:4quark}) and~(\ref{eq:semilep}) give rise to the Feynman
diagrams shown in Fig.~\ref{fig:4FermiFeynman}.  Corrections arising from the electroweak
sector in the SM are known to be subdominant~\cite{Do:2002ky} and are neglected in this
study. They will be included in future work. In the NLO corrections
we neglect the bottom mass dependence. 
Most aspects of this calculation are completely standard. For the one-loop virtual
corrections we use integration-by-parts identities~\cite{Chetyrkin:1981qh} to reduce all
integrals to master integrals. For this calculation only the one-loop
tadpole and one-loop bubble integral with a single massive internal
line are needed, and are trivial to obtain.  Real radiation corrections required for the QCD corrections are obtained from the process $t \to bWg$. These are
straightforward to integrate over the final-state phase
space to obtain the total decay width and helicity fractions. The four-Fermi corrections are infrared finite and only require UV-renormalization.
We regulate all ultraviolet and infrared divergences appearing in
intermediate stages using conventional dimensional regularization
(CDR). The final analytic results are presented in the Appendix. In the
following subsections we focus on technical aspects of the calculation
specific to the SMEFT.

\subsection{Treatment of $\gamma_5$ in SMEFT}

The appearance of $\gamma_5$ in the Feynman rules for this process
indicates that a prescription for handling this quantity when the
space-time dimension $d=4-2\epsilon$ is needed. Different ways of
treating $\gamma_5$ have been extensively discussed in the literature;
for a review see Ref.~\cite{Jegerlehner:2000dz}. In our calculation
$\gamma_5$ appears in three places: through the axial-vector current insertion
$\gamma_{\mu}\gamma_5$; through the axial-tensor current
$\sigma_{\mu\nu}\gamma_5$; through the Dirac structure of each four-Fermi operator.

\subsubsection{An overview of chiral Ward identities}

In the massless limit the SM portion of the Lagrangian of
Eq.~\ref{eq:Lagrangian} has the following chiral U(2) symmetry:
\begin{align}
\label{eq:chiralSU2}
q_L \rightarrow e^{\frac{i}{2}\alpha_i \tau_i} q_L \nonumber\\
q_R \rightarrow e^{-\frac{i}{2}\alpha_i \tau_i} q_R
\end{align}
where $q_{L,R} = (t_{L,R},b_{L,R})$ denote the left-handed and right-handed third-generation quark
doublet.  $\tau_i$ are the generators of this symmetry; $\tau_0$ is
the identity matrix while the $\tau_i$ are the Pauli matrices rescaled
by a factor of $\frac{1}{2}$.
In the SM this symmetry is broken by the quark masses (in this study
we consider only a non-zero top-quark mass).  This leads to well-known
relations between the divergence of the axial-vector current and the
pseudoscalar current as summarized in~\cite{Trueman:1995ca}, and
consequently between correlation functions in the theory.  Any
prescription for $\gamma_5$ in dimensional regularization must satisfy
these relations.

The situation in the SMEFT is slightly more complicated, as not all of
the operators of Section~\ref{sec:SMEFT} satisfy the symmetry of
Eq.~(\ref{eq:chiralSU2}).  Operators such as ${\cal O}_{qq}^{(1)}$ which
are formed from the doublets $q_L$ satisfy the symmetry.  Those
such as ${\cal O}_{qu}^{(1)}$ which feature explict top or bottom
quarks, or those such as ${\cal O}_{tg}$ which couple left-handed
states to right-handed ones, do not.  The standard chiral Ward
identities must be modified in the presence of such operators.

In our calculation we make use of several different schemes for
$\gamma_5$, and check both their consistency with each other and that
they satisfy the appropriate chiral Ward identities.  This gives us
confidence that our treatment of $\gamma_5$ is correct.  We summarize
below what schemes are used in our calculation.  More details appear
in the following subsections.
\begin{itemize}

\item For the QCD-like operators a convenient scheme due to
  Larin~\cite{Larin:1993tq} is available in the literature.  We use
  this approach together with several internal checks to ensure
  correctness of our results.

 \item For the four-Fermi operators we use both a naive anti-commuting
   scheme, which is expected to lead to consistent results in this
   calculation~\cite{Trueman:1995ca}, and the 't
   Hooft-Veltman-Breitenlohner-Maison (HVBM)
   scheme~\cite{Breitenlohner:1975hg}.  We demonstrate that both
   satisfy the chiral Ward identities after appropriate
   renormalization and lead to identical results.
  
\end{itemize}
  
\subsubsection{QCD-like operators: the Larin scheme}

The SM and the QCD-like operators of Eq.~(\ref{eq:dipole}) contain
both the axial-vector current and the axial-tensor current.  The axial-vector current appears in the SM, and its treatment in
dimensional regularization has been studied extensively. A convenient
way to treat the axial vector current is due to Larin~\cite{Larin:1993tq}, and involves
the following replacement:
\begin{equation}
\label{eq:ax_vec}
\gamma_{\mu}\gamma_5 \to \frac{i}{6}\epsilon_{\mu\nu\rho\sigma}\gamma^{\nu}\gamma^{\rho}\gamma^{\sigma}.
\end{equation}
The indices appearing in the Levi-Civita symbol are treated as
$d$-dimensional indices. This replacement violates the chiral  Ward
identities outlined above, leading to the need for an additional finite
renormalization factor:
\begin{equation}
Z_5^{ns} = 1-\frac{ \alpha_s C_F}{\pi} +{\cal O}(\alpha_s^2).
\end{equation}
It is straighforward to check that $\gamma_5$
defined by Eq.~(\ref{eq:ax_vec}) no longer anti-commutes with
$\gamma_{\mu}$ when $\mu$ extends beyond four dimensions. In our
calculation we encounter Dirac traces with either two factors of
$\gamma_5$ or a single $\gamma_5$. In the first
case we can replace the axial-vector current according to
Eq.~(\ref{eq:ax_vec})  immediately at the level of the Feynman rules,
or assume an anti-commuting $\gamma_5$ in order to remove them
completely from the trace, removing the need for $Z_5^{ns}$. We find that both treatments lead to the
same final answer, consistent with the discussion in
Ref.~\cite{Jegerlehner:2000dz}. This serves as a check of our
procedure.  We also reproduce exactly the known SM QCD results for the
total width and helicity fractions.

The axial-tensor current does not appear in the Standard Model Feynman
rules. We note that in $d=4$, the Chisholm identity can be used to
rewrite the axial-tensor current according to
\begin{equation}
\sigma_{\mu\nu}\gamma_5 = -\frac{i}{2}\epsilon_{\mu\nu\rho\sigma} \sigma^{\rho\sigma}\footnote{The conventions here are such that $\gamma_5 = i \gamma^0\gamma^1\gamma^2\gamma^3$ and $\epsilon^{0123} = +1$}.
\end{equation}
By avoiding the introduction of a non-anticommuting $\gamma_5$ all
Ward identities are preserved, indicating that no additional finite
renormalization is needed for the axial-tensor current, unlike for the axial-vector current. This
observation has also been made in previous studies in heavy-quark
effective theory~\cite{Broadhurst:1994se}. To check this result we
have also used the 't Hooft-Veltman replacement
\begin{equation}
\gamma_5 \to \frac{i}{24}\epsilon_{\mu\nu\rho\sigma}\gamma^{\mu} \gamma^{\nu}\gamma^{\rho}\gamma^{\sigma}
\end{equation}
and have found the same final result as obtained with the Chisholm
replacement. We note that the Chisholm replacement is computationally
more efficient, as it leads to fewer $\gamma$-matrices within Dirac
traces.

To have an independent check of the validity of the Larin scheme in the
presence of SMEFT operators we verify the corresponding chiral Ward
identities through explicit calculation.  We consider ${\cal
  O}_{tg}$ as an example.  This operator violates the chiral U(2)
symmetry since it contains a current coupling a left-handed and
right-handed state.  A variation of the Lagrangian under the symmetry
transformation considered leads to the following relation between
correlation functions:
\begin{align}
\braket{\partial_\mu(\overline{t}(x)\gamma^\mu\gamma_5t(x))t(x_1)\overline{t}(x_2)} =\;& 2im_t \braket{\overline{t}(x)\gamma_5t(x)t(x_1)\overline{t}(x_2)}\nonumber\\
&-i\gamma_5\braket{t(x)\overline{t}(x_2)}\delta^d(x-x_1)-i\braket{t(x_1)\overline{t}(x)}\gamma_5\delta^d(x-x_2) \nonumber\\
&-\sqrt{2}i\frac{v}{\Lambda^2}C_{tg} \braket{\left(\overline{t}(x)\sigma^{\mu\nu}\gamma_5T^At(x)G^A_{\mu\nu}\right)t(x_1)\overline{t}(x_2)}.
\end{align}
The derivation implicitly assumes that the functional measure
transforms trivially under the chiral rotation, i.e. the symmetry is
non-anomalous. This assumption is supported by discussions
in~\cite{Donoghue:1992dd}, which indicate that any anomalous terms are
proportional to the square of the Wilson coefficients.  The left- and right-handed side of the identity agree with each other in the Larin scheme described above, giving us confidence in our results for the decay widths calculated in SMEFT.

Since our top decay calculation does not involve traces over triple
axial vector current insertions that appear in triangle fermion loops,
we can extend the indices of Levi-Civita symbol safely from $4$ to $d$
dimensions. We have verified through explicit calculation that all
differences in this treatment of the Levi-Civita symbol in combination with both the Chisholm replacement as well as the 't Hooft-Veltman replacement appear only at $\mathcal{O}(\epsilon)$ for our observables.

\subsubsection{Four-Fermi operators: naive anticommuting and HVBM schemes}
In the case of diagrams involving four-Fermi interactions we follow a two-pronged approach to obtain consistent results.
First we employ the {\sc FeynCalc}~\cite{Shtabovenko:2016olh} internal scheme in which an anti-commuting $\gamma_5$ is assumed in combination with
\begin{align} 
\mathrm{Tr}[\gamma_5 \gamma^\mu\gamma^\nu\gamma^\rho\gamma^\sigma] = 4i\epsilon^{\mu\nu\rho\sigma},
\end{align}
where the Levi-Civita symbol is treated as a strictly $4$-dimensional
object.  Combining an anti-commuting $\gamma_5$ with a non-vanishing
trace as done here is strictly speaking inconsistent but is known to
lead to correct results in the case of one-loop corrections~\cite{Gamma5tests}. Since there exists no formal proof for this statement we
verify our results again by confirming that the scheme preserves the
Ward identities between correlation functions associated with
Eq.~\ref{eq:chiralSU2}, which in the presence of the four-Fermi operators reads
\begin{align}
\braket{\partial_\mu(\overline{t}(x)\gamma^\mu\gamma_5t(x))t(x_1)\overline{t}(x_2)} =\;& 2im_t \braket{\overline{t}(x)\gamma_5t(x)t(x_1)\overline{t}(x_2)}-i\gamma_5\braket{t(x)\overline{t}(x_2)}\delta^d(x-x_1)\nonumber\\
&-i\braket{t(x_1)\overline{t}(x)}\gamma_5\delta^d(x-x_2),
\end{align}
and analogously for the bottom quark current. We confirm through
explicit calculation that the naive anti-commuting scheme, as
implemented in {\sc FeynCalc}, preserves the Ward identity in the presence
of four-Fermi operators that conserve the chiral U(2) symmetry.  As
mentioned previously some operators (e.g. $C_{qu}^{(1)}$) explicitly
violate the chiral transformation, and consequently satisfy a more complicated identity.
As a second check we employ the self-consistent 't
Hooft-Veltman-Breitenlohner-Maison (HVBM)
scheme~\cite{Breitenlohner:1975hg}, as it is implemented in {\sc
  TRACER}~\cite{Jamin:1991dp}.  This involves splitting all $d$-dimensional objects into sums of their $4$-dimensional parts (denoted by a bar) and $d-4$-dimensional (denoted by a hat) parts:
\begin{align}
g^{\mu\nu} = \overline{g}^{\mu\nu} + \hat{g}^{\mu\nu} \;\;\;\;\;\;\;\; q^\mu = \overline{q}^\mu + \hat{q}^\mu,
\end{align}
where external momenta and the Levi-Civita symbol are treated as
purely $4$-dimensional. We note that each of the bared and hatted
objects acts as a projector for the $4$ and $d-4$-dimensional
subspaces respectively.  The results obtained in this approach violate
the chiral Ward identities which need to be restored through the
introduction of finite corrections stemming from evanescent operators,
as described in~\cite{Adel:1994my,Herrlich:1994kh,Trueman:1995ca}.
After the inclusion of these finite corrections the results obtained
with this approach must agree through ${\cal O} (\epsilon^0)$
with those obtained using {\sc FeynCalc}.  We have checked for several
operators under consideration that this is indeed the case.

\subsection{Ultraviolet renormalization}
The ultraviolet (UV) renormalization of the external states is
performed in the on-shell scheme, similar to the renormalization
usually performed in the Standard Model. The only non-vanishing terms
stem from the QCD corrections to the external top-quark line. The
quark self-energy corrections from the four-Fermi operators are
independent of the their respective momenta and therefore do not change
the wave function renormalization. Neither electric charge, weak
mixing angle nor $W$-wave function receive any contributions, since
they exclusively depend on the gauge-boson self energies.

To calculate the gluonic contribution to the wave function renormalization we note that the quark self energy can be decomposed in SMEFT in the same way as in the SM:
\begin{align}
\Sigma_q(p^2) = \slashed{p}P_L \Sigma^L_q(p^2) + \slashed{p}P_R \Sigma^R_q(p^2) + m_t \Sigma^S_q(p^2),
\end{align} 
with the chirality projection operators $P_{\nicefrac{R}{L}} =
\frac{1}{2}(1\pm\gamma_5)$.  We therefore can calculate the left and right-handed quark field renormalizations $\delta Z^{\nicefrac{L}{R}}_q$ from the quark self energy according to
\begin{align}
\delta Z^{\nicefrac{R}{L}}_{q} = -\Sigma^{\nicefrac{R}{L}}_q(m_t^2) - m_q^2 \frac{\partial}{\partial p^2}\textrm{Re}\Big\{\Sigma^L_q(p^2)+\Sigma^R_q(p^2)+2\Sigma^S_q(p^2)\Big\}\bigg|_{p^2=m_q^2}.
\end{align}
Calculating this expression in the SMEFT for the top quark yields
\begin{align}
\delta Z^{\nicefrac{R}{L}}_{t} = \frac{C_F g_s}{32\pi^2}\frac{d-1}{d-3}\frac{A_0(m_t^2)}{m_t^2}\bigg [g_s (d-2) - 2\sqrt{2} x_vm_t^2C_{tg} \bigg],
\end{align}
Our conventions are such that the tadpole master integral is 
\begin{align}
A_0(m^2) = m^2 \left[\frac{1}{\epsilon}+1-\log\left(\frac{m^2}{\mu^2}\right)\right]
\end{align}
with renormalization scale $\mu$, which will in the end be set to the
top mass $m_t$ in our numerical studies.  This also explicitly
confirms that left and right-handed top quarks still receive the same
contributions in SMEFT QCD. Furthermore, the corresponding field
renormalizations of the bottom quark vanish identically in the limit
$m_b=0$.

It is necessary to introduce additional counterterms by renormalizing the SMEFT Wilson coefficients themselves. This is customarily done in \msbar~\cite{Jenkins:2013zja,Jenkins:2013wua,Alonso:2013hga} and can be achieved in the QCD sector through the replacement
\begin{align}
C_{tW} \rightarrow C_{tW} + \frac{C_F g_s^2}{16\pi^2\epsilon}C_{tW} - \frac{C_F \bar{g}g_s}{16\pi^2\epsilon} C_{tg}\nonumber\\
C_{bW} \rightarrow C_{bW} + \frac{C_F g_s^2}{16\pi^2\epsilon}C_{bW} - \frac{C_F \bar{g}g_s}{16\pi^2\epsilon} C_{bg},
\end{align}
introducing further operator mixing.  In the case of the four-Fermi
operators we renormalize the operator $C_{\phi q}^{(3)}$ by shifting the $CKM$ matrix as mentioned before. We choose
\begin{align}
\delta C_{\phi q}^{(3)} = \frac{\overline{g}^2C^{(3)}_{lq}}{48\pi^2\epsilon},
\end{align}
for a lepton pair $l, \nu_l$ in the loop, as well as
\begin{align}
\delta C_{\phi q}^{(3)} = \frac{\overline{g}^2-3y_t^2}{48\pi^2\epsilon}\left(C_{qq}^{(1)}+(2N_C-1)C_{qq}^{(3)}\right),
\end{align}
for the $t,b$ loop, where the non-vanishing top mass in the loop gives rise to the top Yukawa $y_t = \frac{\sqrt{2}m_t}{v}$. Correspondingly we find
\begin{align}
\delta C_{\phi q}^{(3)} = \frac{\overline{g}^2}{48\pi^2\epsilon}\left(C_{qq,\textrm{light}}^{(1)}+(2N_C-1)C_{qq,\textrm{light}}^{(3)}\right),
\end{align}
for light quarks in the loop. We report the counterterms here for
completeness but omit the light quark loop from our analysis, since the associated Dirac structure could only be achieved by integrating out a heavy neutral vector boson that changes quark flavor from a UV completion.
The counterterms found here are in agreement with the ones reported
in~\cite{Alonso:2013hga,Jenkins:2013wua}. We note  that for
consistency $\alpha_s$ is run from the $Z$-scale up to the top mass scale utilizing the two loop SM-running found in the literature~\cite{Caswell:1974gg}.

\section{Numerical results}
\label{sec:numerics}

We present our numerical results in this section. We assume
$\Lambda = 500$ GeV throughout this section, which makes the Wilson
coefficients under discussion dimensionless. The input parameters are
summarized in Table~\ref{tab:input}.  The measured values of the top
decay width and helicity fractions we use to constrain the operators
are taken from the PDG~\cite{Tanabashi:2018oca}:
\begin{gather}
\Gamma_\textrm{tot}^{\rm exp} = 1.41^{+0.19}_{-0.15}\; \textrm{GeV},\;\;\;\; F^\textrm{exp}_{L} = 0.687 \pm 0.018, \;\;\;\; F^\textrm{exp}_{-} = 0.320 \pm 0.013.
\end{gather}
We study projections for higher integrated luminosities relevant for a
high-luminosity LHC (HL-LHC) and a potential future $e^+e^-$ collider
later in this section. Since we perform a fit to only a limited set of observables, rather than a global fit such as considered in~\cite{Buckley:2015lku,AguilarSaavedra:2018nen,Hartland:2019bjb}, our numerical results should only be considered representative of the achievable bounds on the studied operators. 

\begin{table}[h!]
\centering
\begin{tabular}{|cc||cc|}
\hline\hline  
 $M_Z$ & $91.1876\,\textrm{GeV}$ & $M_W$ & $80.379\,\textrm{GeV}$  \\ 
 
 $v$& $246 \,\textrm{GeV}$ & $m_t$ & $173.0 \,\textrm{GeV} $ \\ 
 
 $m_b$& $4.78  \,\textrm{GeV}$ & $G_F$ & $1.1664 \times 10^{-5} \,\textrm{GeV}^{-2}$  \\ 

 $\alpha^{-1}_\textrm{em}$ & $137.036$ & $\alpha_s(M_Z)$ & $0.1185$ \\ 

\hline\hline
\end{tabular}
\mycaption{Input parameters for the calculation, taken from~\cite{Tanabashi:2018oca}. The value of the $SU(2)$ coupling $\bar{g}$ is calculated from the Fermi constant $G_F$ and the fine structure constant $\alpha_\textrm{em}$.\label{tab:input}}
\end{table}

\subsection{QCD operators}
We begin by discussing the contributions from QCD-like operators,
namely $C_{tW}, C_{bW}$, $C_{\phi tb}$, $C_{tg}$, and $C_{bg}$. A
similar analysis of these operators was performed in
Ref.~\cite{Zhang:2014rja}\footnote{We find an identical analytic expressions for the total width $\Gamma_\textrm{tot}$ and longitudinal helicity fraction $F_L$ to leading order in $x_b$. We find however a different dependence on $C_{tg}$ in the case of the transverse helicity fractions $F_\pm$.}, focusing however only on the constraints
derived for $C_{tW}$ and $C_{tg}$. We update the constraints on these
operators and discuss constraints on the remaining ones.  At LO, the total width is only a function of $C_{tW}, C_{bW}$, and $C_{\phi tb}$, which enter through the W-vertex. The NLO corrections induce sensitivity to $C_{tg}$ comparable to that of $C_{\phi tb}$. We note that the total width is independent of $C_{bg}$, due to the operator being helicity suppressed, as evident from the analytic expression of Eq.~(\ref{app:width}). We find that the total width is significantly more constraining for $C_{tW}$ than $C_{bW}$, and that the constraints on these two operators are both stronger than the bounds on $C_{\phi t b}$ and $C_{tg}$.  The Wilson coefficients are also constrained through the longitudinal, positive transverse and negative transverse helicity fractions.  The longitudinal rate $F_L$ is again independent of $C_{bg}$, as seen from Eq.~(\ref{app:fl}). We note that the variation of $C_{bW}$ significantly alters the positive transverse helicity fraction, $F_{+}$.

In order to derive constraints on the Wilson coefficients from the
current experimental measurements of both the total width and the
helicity fractions, we perform a one-parameter $\chi^2$ fit for each
Wilson coefficient by keeping only one of them nonzero at a time.  We
also report projections for bounds potentially obtainable at a high
luminosity LHC (HL-LHC) after collecting 3 ab$^{-1}$ of data. The $\chi^2$ function is defined through 
\begin{equation}
\Delta \chi^2=\sum_{ij}(O_i^{\rm theo}-O_i^{\rm exp})(\sigma^2)^{-1}_{ij}(O_j^{\rm theo}-O_j^{\rm exp})\, ,
\end{equation}
where $O_i^{\rm exp}$ are the measured observables ($\Gamma_{\rm tot}$, $F_L$, and $F_-$),
$O_i^{\rm theo}$ are their predicted values in the SMEFT and
$\sigma^2_{ij}=\sigma_i\rho_{ij}\sigma_j$, where
$\sigma_i$ are the uncertainties and $\rho$ is the correlation matrix,
\begin{equation}
\rho=\left(
\begin{array}{ccc}
1.0 & 0 & 0    \\
0 & 1.0& -0.87  \\
0 & -0.87 & 1.0  \\
\end{array}
\label{cormat}
\right)\, .
\end{equation}
The correlation matrix comes from a CMS measurement of the helicity
fractions~\cite{Khachatryan:2016fky}.  We assume that it is applicable to the PDG average and
that the total width is uncorrelated with the $F_i$ measurements.
We believe that these simple assumptions capture the features of a more
complete analysis.  For the asymmetric errors in $\Gamma_{\rm  tot}$, we combine them in quadrature. i.e. $\sigma_{\Gamma_{\rm tot}}= \sqrt{(\sigma_{\Gamma_{\rm tot}}^{\rm upper})^2+(\sigma_{\Gamma_{\rm tot}}^{\rm lower})^2}$. 
\begin{table}[h]
\centering
\begin{tabular}{c||c|c|c|c}
 & Current     & HL-LHC ($f_{\rm syst}=1/2$) & HL-LHC ($f_{\rm syst}={1\over \sqrt{N}}$) \\
\hline\hline
$C_{tW}$      & $0.06 \pm 0.08$    & $0.06 \pm0.03$      & $0.06 \pm 0.01$   \\

$C_{tg}$       &$-4.25 \pm6.42$     & $-4.52 \pm 2.34$    & $ -4.25 \pm 0.52$  \\

$C_{bW}$     &$0.80 \pm0.89$      & $ 0.80 \pm0.32$     & $0.80 \pm 0.07$   \\

$C_{bg}$      & $-13.54\pm13.49$ & $-13.54 \pm 4.83$  & $-13.54 \pm 1.10$ \\

$C_{\phi tb}$&$4.35 \pm 5.98$     & $4.47 \pm 2.16$      & $4.35 \pm 0.49$    \\
\hline\hline
\end{tabular}
\mycaption{Best for the Wilson coefficients of the QCD operators with their respective errors at 68\% CL. The scale $\Lambda$ is assumed to be 500 GeV.
The first column shows the results based on the current LHC data with the luminosity of 20 fb$^{-1}$ and the rest of them the projection based on the HL-LHC with the luminosity of 3 ab$^{-1}$. For the projection of the uncertainties at HL-LHC, the statistical uncertainties scale like $1\over \sqrt{N}$ while the systematic uncertainties are scaled by a factor of $f_{\rm syst}$. 
\label{tab:xisqfittable-QCD}}
\end{table}

\begin{figure}[h!]
\centering
\includegraphics[width=.47\textwidth]{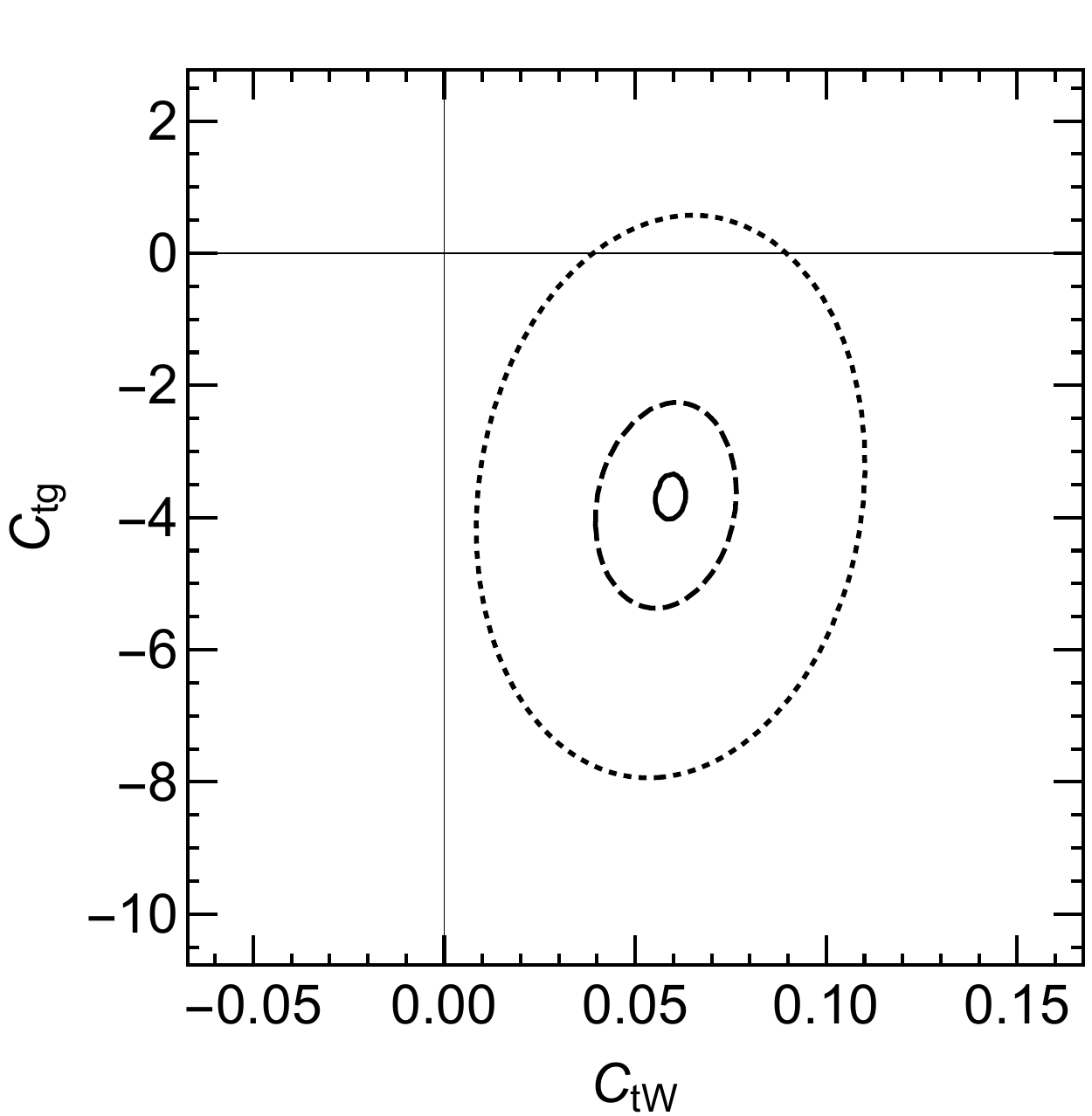}
\includegraphics[width=.489\textwidth]{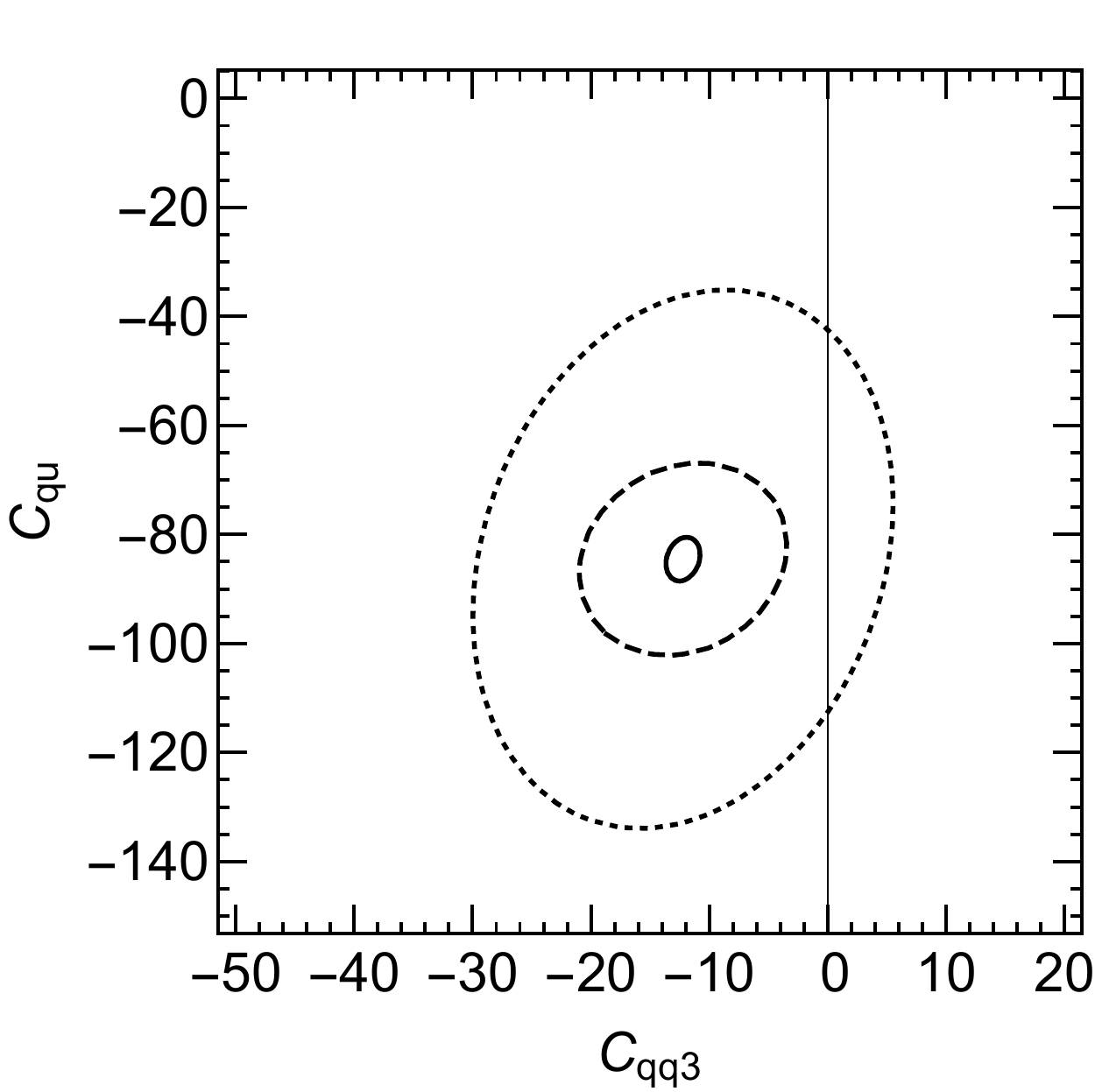}
\mycaption{Two-dimension $\chi^2$ fit at 95\% CL for the QCD
  operators, $C_{tW}$ and $C_{tg}$ (left), and  the four-fermion
  operators, $C^{(3)}_{qq}$ and $C_{qu}$ (right).
  We assume $\Lambda = 500\textrm{GeV}$ and show the dimensionless coefficient. The dotted contour corresponds to the current measurement, while the dashed and solid contours correspond to $f_{\rm syst}=1/2$ and  $f_{\rm syst}=1/\sqrt{N}$ at the HL-LHC.
\label{fig:2d}}
\end{figure}

The results of the 1-parameter fits are summarized in Table~\ref{tab:xisqfittable-QCD}.  The first column of Table~\ref{tab:xisqfittable-QCD} shows the results
based on the current LHC data with the luminosity of 20 fb$^{-1}$.
The other columns show projections based on the HL-LHC with a luminosity of 3 ab$^{-1}$.
For the projections  we reduce the statistical errors as $1/\sqrt{N}$,
where the number of events $N$ scales like the integrated luminosity.
We consider two assumptions for the scaling factor associated with the
systematic error, $f_{\rm syst}$. 
\begin{enumerate}

\item $f_{\rm syst}=1/2$: this is close to a recommendation proposed
  by ATLAS  where all the systematic errors are scaled by a factor of
  1/2 \cite{atlas-note,Azzi:2019yne}.

 \item  $f_{\rm syst}=1/\sqrt{N}$: this is based on a CMS proposal
   used in previous projections~\cite{cms-note}, where the systematic
   error is assumed to scale like the statistical uncertainty.  This
   is the more optimistic of the two scenarios.
  
\end{enumerate}  
We find that in the second projection that the bounds on the QCD operators can
be tightened by at least an order of magnitude at the HL-LHC, while in
the first projection the uncertainty reduction is less.  
Both $C_{tW}$ and $C_{bW}$ are already significantly constrained with the current
measurements.  We have also performed a two-dimensional $\chi^2$ fit at
95\% CL for $C_{tW}$ and $C_{tg}$ as shown in the left panel of
Fig.~\ref{fig:2d} to study potential correlations between these
parameters. The dotted contour corresponds to the current measurement,
while the dashed and solid contours correspond to $f_{\rm syst}=1/2$
and  $f_{\rm syst}=1/\sqrt{N}$ at the HL-LHC.  Only a weak correlation
is observed. 

Previous constraints on these EFT operators at leading order using
top-quark observables can be found in the
literature~\cite{Buckley:2015lku,Zhang:2010dr}. We have checked that
when our calculation is truncated at LO the bounds we find agree
with those previously obtained.  A more complete analysis would
include the NLO electroweak corrections in the Standard
Model~\cite{Do:2002ky}.  These are outside the scope of the simple fit
presented here.  An important point learned from the
above table is that at a HL-LHC, the bounds on the loop-induced Wilson
coefficients $C_{tg}$ and $C_{bg}$ can approach unity.  This demonstrates that higher-order effects in the SMEFT can be significantly probed during the future LHC program.

\subsection{Four-fermion operators}

\begin{figure}[htbp]
\centering
\includegraphics[width=.49\textwidth]{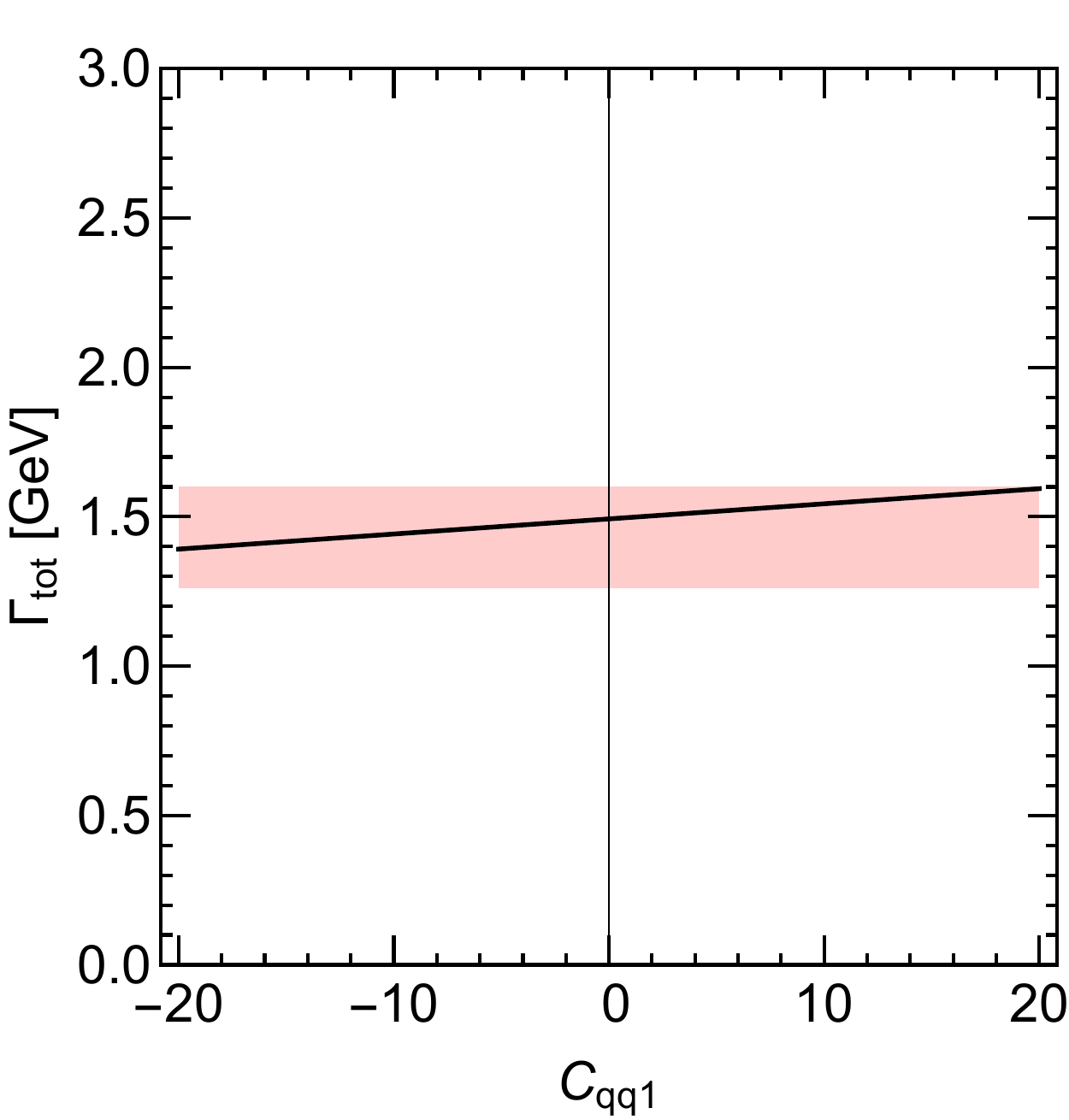}
\includegraphics[width=.49\textwidth]{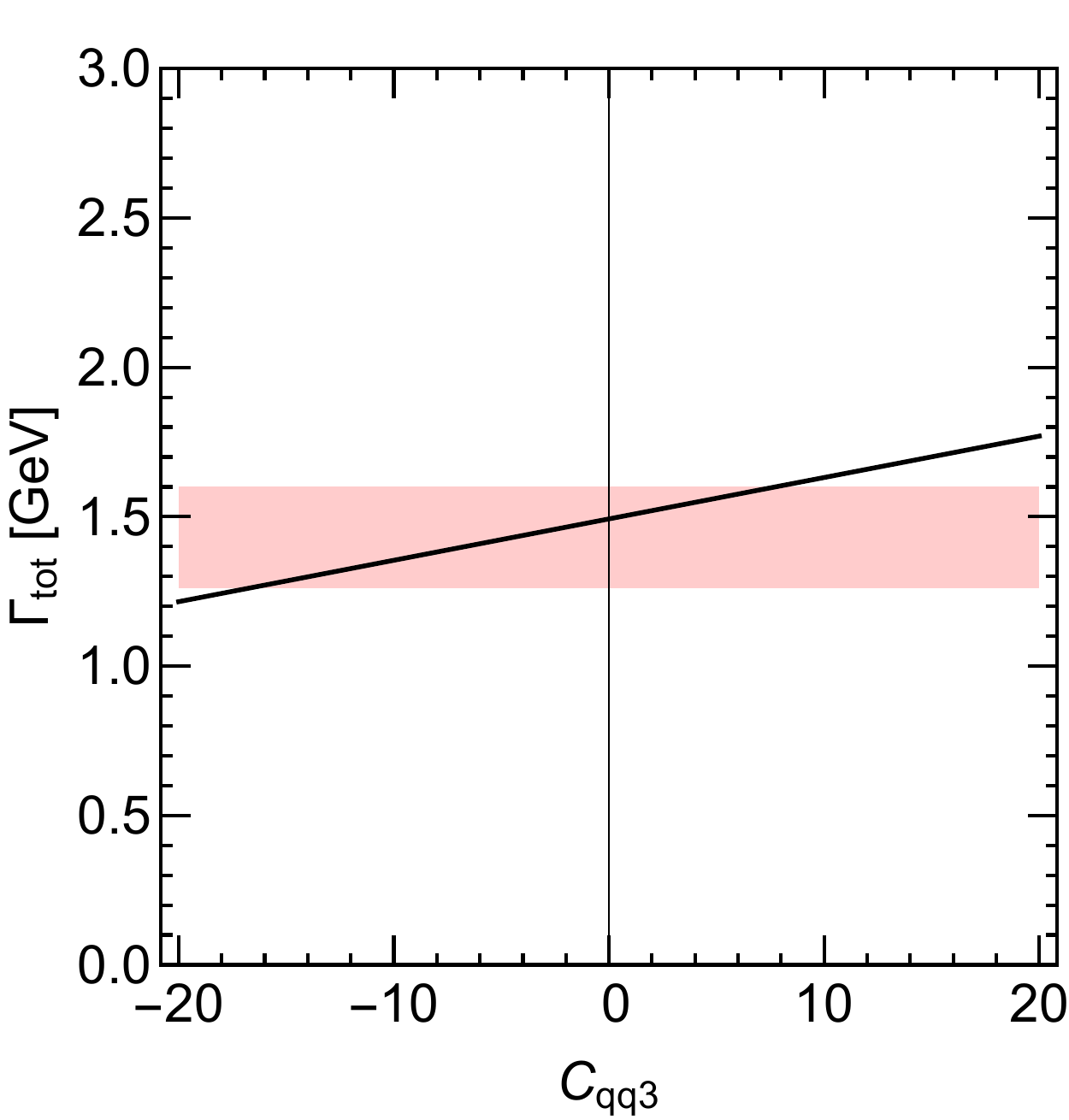}
\includegraphics[width=.49\textwidth]{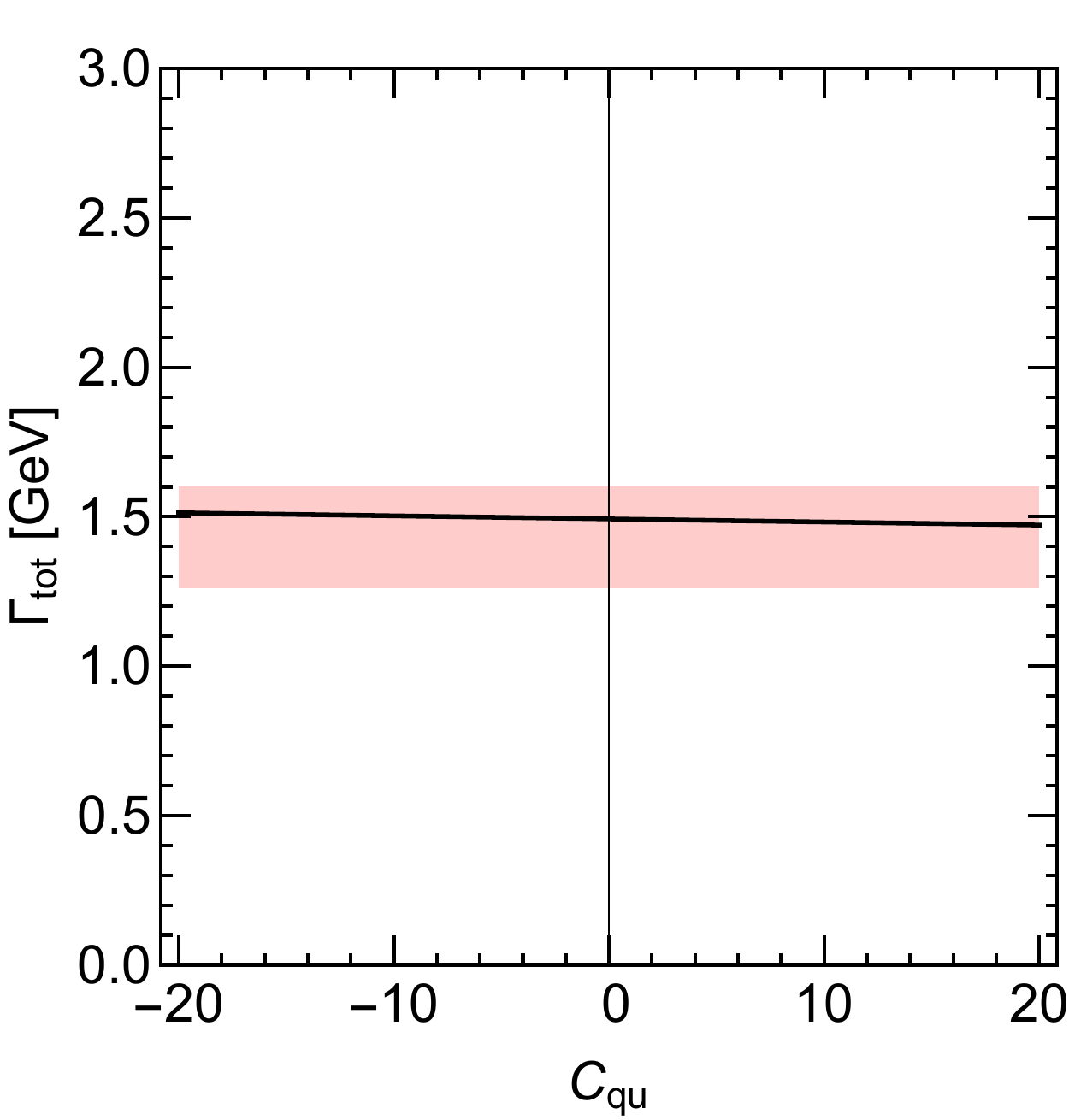}
\includegraphics[width=.49\textwidth]{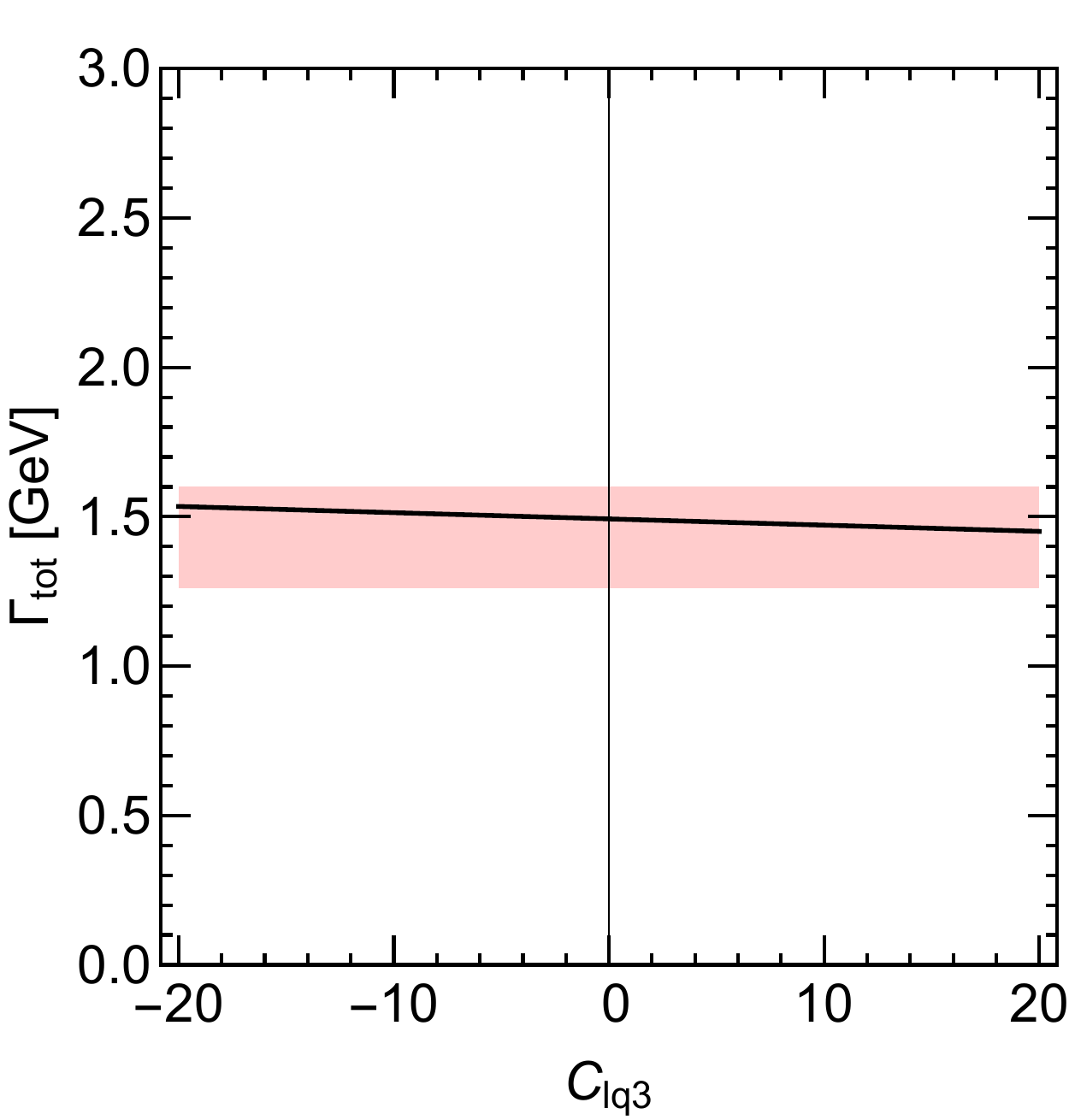}
\vspace{-2ex}
\mycaption{Comparison of the Wilson coefficient sensitivity of the total decay width $\Gamma_\textrm{tot}$ for the four-Fermion operators. We assume $\Lambda = 500\textrm{GeV}$ and show the dimensionless coefficient. The shaded band shows the $1\sigma$ region around the experimental decay width.
\label{fig:TotWidthPlots}}
\end{figure}

We next present and discuss the bounds on the four-Fermi operators to
which we are sensitive: $C^{(1)}_{qu}$, $C^{(8)}_{qu}$,
$C^{(1)}_{qq}$, $C^{(3)}_{qq}$, and $C^{(3)}_{lq}$. The sensitivity of the total width to the
different Wilson coefficients is shown in
Fig.~\ref{fig:TotWidthPlots}.  Since the observables we consider are
only sensitive to the combination
$C_{qu}=C^{(1)}_{qu}+{4\over3}C^{(8)}_{qu}$ we plot only that
structure.  The shaded band represents the $1\sigma$ region around the experimentally measured value, while the solid black line is the NLO result as a function of a single Wilson coefficient. We find with the current experimental errors that the total width is only weakly sensitive to these operators, with the exception of $C^{(3)}_{qq}$. We also find that $C^{(3)}_{lq}$ only appears in the total width and drops out from the helicity fractions after an expansion in 1/$\Lambda^2$. $C^{(3)}_{lq}$ is however only weakly bounded by the total width as is evident from Fig.~\ref{fig:TotWidthPlots}. 

\begin{figure}[htbp]
\centering
\includegraphics[width=.49\textwidth]{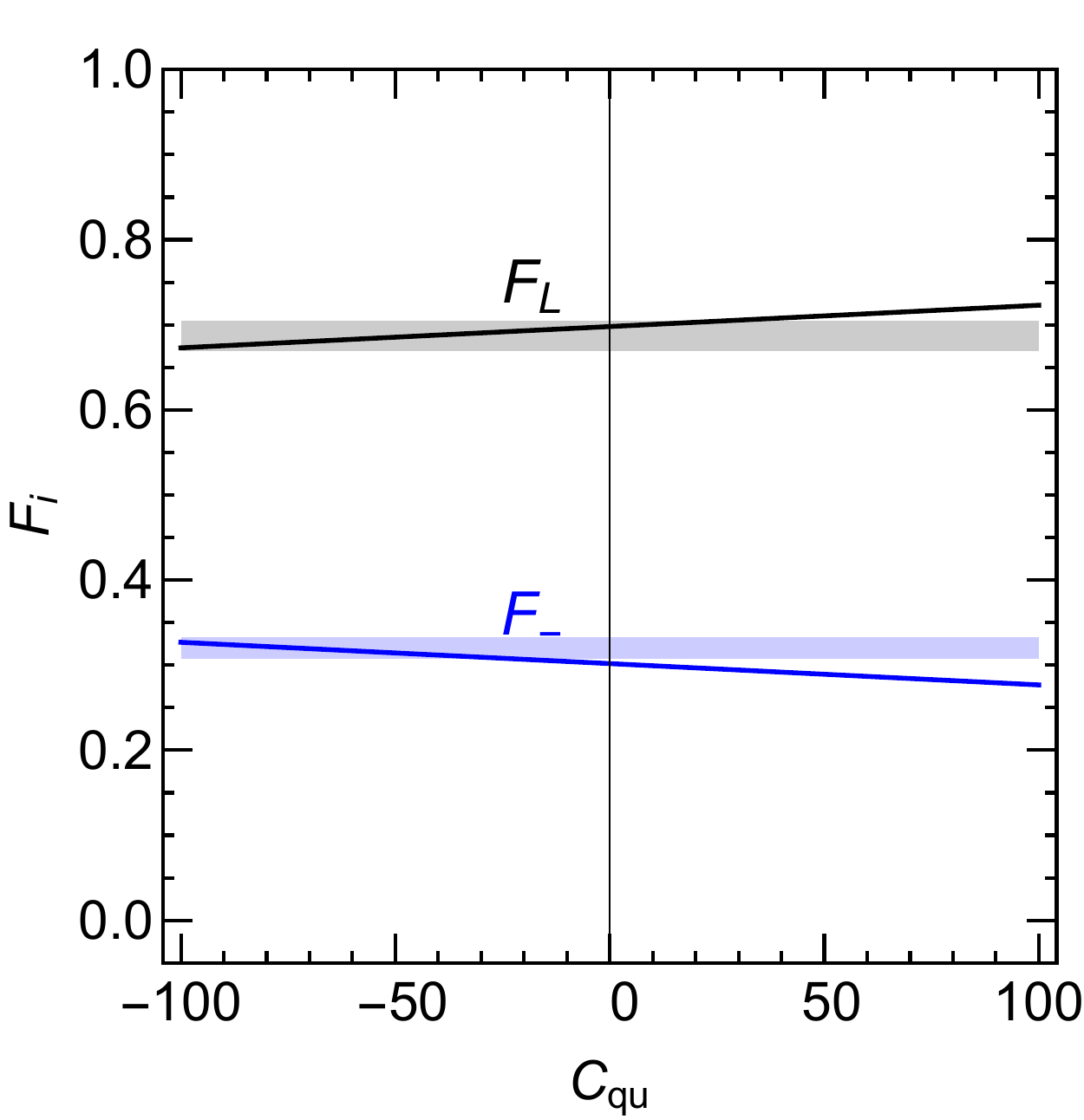}
\mycaption{Comparison of the Wilson coefficient sensitivity of the longitudinal and negative transversal helicity fractions for the four-Fermion operators. We assume $\Lambda = 500\textrm{GeV}$ and show the dimensionless coefficient. The shaded band shows the $1\sigma$ region around the experimentally measured fractions. 
\label{fig:FractionsPlots}}
\end{figure}

As before there are additional constraints set by the helicity
fractions, $F_L$ and $F_-$. Fig.~\ref{fig:FractionsPlots} shows the
helicity fractions as functions of $C_{qu}$. The shaded band is
again the $1\sigma$ region around the experimentally measured
fractions. The solid black lines show the functional dependence of the
helicity fractions on each Wilson coefficient at NLO. We see that the
results are quite different than those observed for the total width.
$C_{qu}$ is now probed by $F_L$, and $F_-$, but we lose
sensitivity to all other four-fermion operators.  We note that no observables are sensitive to $C^{(1)}_{qq}$ and $C^{(3)}_{lq}$.

The global constraints on the Wilson
coefficients are derived through a one-parameter $\chi^2$ fit for each
Wilson coefficient. The resulting best fits and corresponding bounds
are summarized in Table~\ref{tab:xisqfittable-4fermi}. We again
compare the 68\% CL bounds derived from current LHC data with the projected ones for HL-LHC at 3 ab$^{-1}$.
As expected $C^{(3)}_{qq}$ is constrained most strongly.  This bound
is mainly set by the total width. 
The bounds on $C_{qu}$ are around a factor of two weaker in comparison,
stemming from $F_L$ and $F_-$. $C^{(1)}_{qq}$ and
$C^{(3)}_{lq}$ are not constrained through any of the observables.  We
have also performed a two-dimension $\chi^2$ fit at 95\% CL for
$C^{(3)}_{qq}$ and $C_{qu}$
as shown in the right panel of Fig~\ref{fig:2d}. The dotted contour
corresponds to the current measurement, while the dashed and solid
contours correspond to $f_{\rm syst}=1/2$ and  $f_{\rm
  syst}=1/\sqrt{N}$ at the HL-LHC.  These bounds are complementary to
the ones in the literature derived from direct production of four
final-state heavy flavors~\cite{Azzi:2019yne}.

We note that the bounds on all four-Fermi Wilson coefficients are very weak with
the current data. In fact, if we estimate the energy scale probed by
each observable as $\Lambda/\sqrt{C_X}$, we find that the
currently-accessible energy scales are less than the top-quark mass.
This indicates that the EFT expansion is not compatible with the
current experimental errors. At an HL-LHC, the bounds on all the
four-Fermi operators improve significantly.  In particular,
constraints on the Wilson coefficients $C_{qq}^{(3)}$, $C_{qu}^{(1)}$
and $C_{qu}^{(8)}$ approach unity and the effective energy scale
probed is significantly above the top-quark mass, indicating that
these higher-order effects can be meaningfully probed during the
future LHC program.
\begin{table}[ht]
\centering
\begin{tabular}{c||c|c|c|c}
 & Current & HL-LHC ($f_{\rm syst}=1/2$)  & HL-LHC ($f_{\rm syst}={1\over \sqrt{N}}$)  \\
\hline\hline
$C^{(1)}_{qq}$   & $43.76 \pm 128.16$    & $ 43.76 \pm 64.08$   & $ 43.76 \pm 10.46$   \\

$C^{(3)}_{qq}$   &$-5.97 \pm 17.47$        & $ -5.97 \pm 8.74$    & $ -5.97 \pm 1.43$     \\

$C_{qu}$           &$-77.51 \pm 73.46$      & $ -80.85 \pm 26.59$   & $-77.51 \pm 6.00$    \\

$C^{(3)}_{lq}$    &$39.57 \pm 115.90$     & $ 39.57 \pm 57.95$    & $39.57 \pm 9.46$  \\
\hline\hline
\end{tabular}
\mycaption{Best $\chi^2$ fit for the Wilson coefficients of the four-Fermion operators with their respective errors at 68\% CL. The scale $\Lambda$ is assumed to be 500 GeV.
The second column shows the results based on the current LHC data with the luminosity of 20 fb$^{-1}$ and the rest of them the projection based on the HL-LHC with the luminosity of 3 ab$^{-1}$. For the projection of the uncertainties at HL-LHC, the statistical uncertainties scale like $1\over \sqrt{N}$ while the systematic uncertainties are scaled by a factor of $f_{\rm syst}$. 
\label{tab:xisqfittable-4fermi}}
\end{table}

\begin{figure}[h!]
\centering
\includegraphics[width=0.5\textwidth]{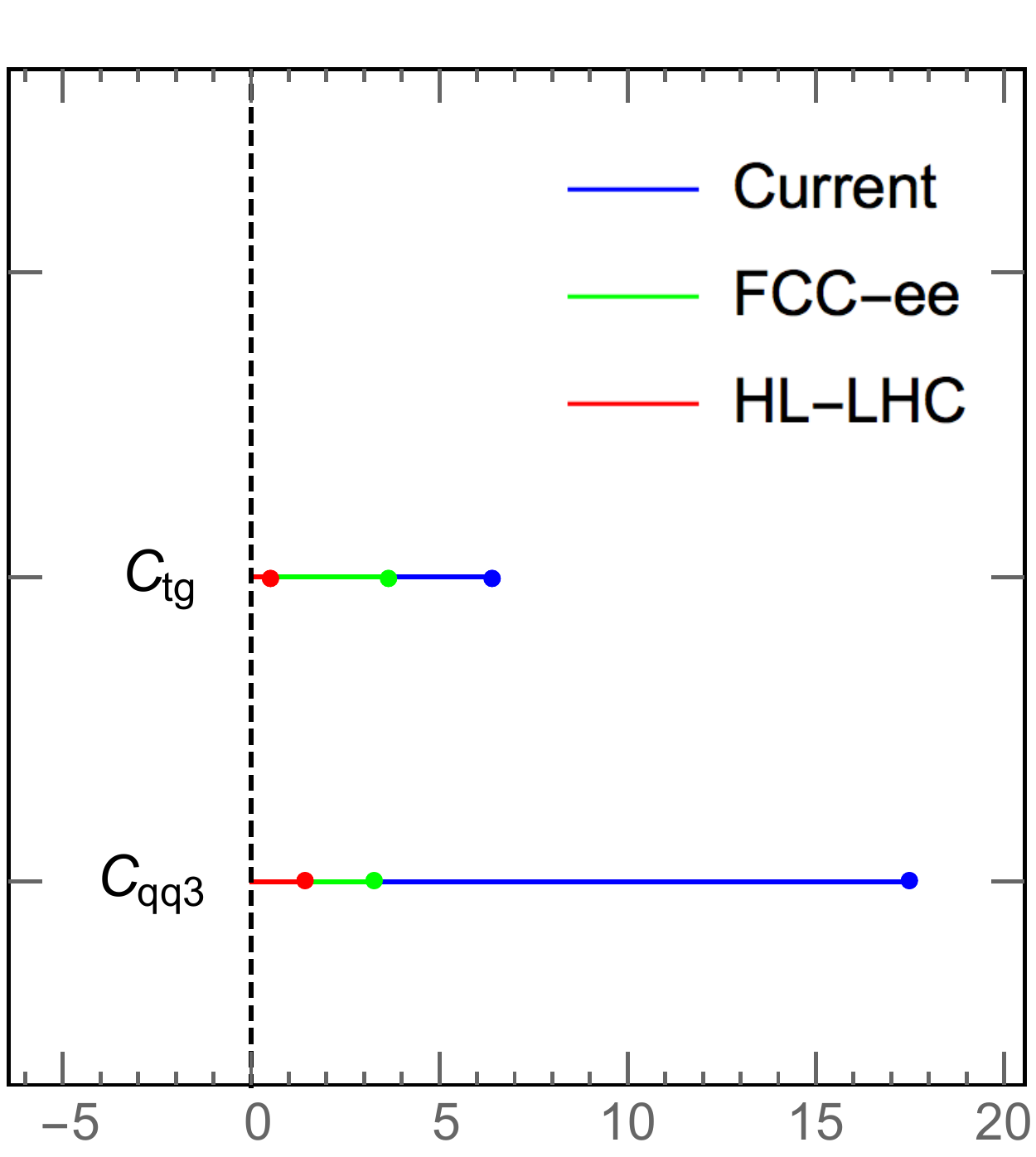}
\mycaption{68\% CL uncertainties for the QCD operator, $C_{tg}$, and four-fermi operator, $C^{(3)}_{qq}$. 
The blue, green, and red bars correspond to the uncertainties from the current measurement, FCC-ee, and HL-LHC ($f_{\rm syst}=1/\sqrt{N}$), respectively. 
\label{fig:barplot}}
\end{figure}

Finally, we study as well projected errors for a potential future
FCC-ee $e^+e^-$ machine.  Details of this project are provided
in~\cite{Benedikt:2018qee}, where it is indicated that the top-quark
width can be probed with a precision of 45 MeV.  We use this estimated
error together with the more optimistic HL-LHC systematic error
estimate to check what bounds the total width can provide on several
example operators for each machine.  The 68\% CL uncertainties for the
current measurement and the future colliders for $C^{(3)}_{qq}$ and
$C_{tg}$ are shown in Fig.~\ref{fig:barplot},
where the blue, green, and red bars correspond to the uncertainties from the current measurement, FCC-ee, and HL-LHC ($f_{\rm syst}=1/\sqrt{N}$), respectively. We find that the bounds are significantly improved at both future colliders.


\section{Conclusions}
\label{sec:conc}

In this manuscript we have studied the next-to-leading order
corrections to top quark decays within the SMEFT.  Our calculation
includes a more complete set of operators than previously studied, in
particularly all contributing four-Fermi operators.  We have two primary motivations.
First, this work is a step toward a complete calculation of NLO effects
within SMEFT, which we believe will eventually be required by the
experimental uncertainties.  Second, this work tests the question of
whether loop-induced operators can be probed with either current or
potential future collider data.

We have addressed technical aspects associated with higher-order
calculations in the SMEFT containing $\gamma_5$.  Chiral Ward identities
play an important role in imposing consistency of $\gamma_5$
prescriptions at higher-order in dimensional regularization, which we
demonstrate by checking the consistency of several different schemes
at one-loop order.  Finally, we have presented numerical bounds on the
considered operators given current and projected future
uncertainties. We find that future machines such as
the HL-LHC or a future $e^+e^-$ collider can provide important
constraints on SMEFT operators that first appear at higher-orders.


\section*{Acknowledgments}
We thank C. Zhang for useful discussions.  R.~B. is supported by the DOE contract DE-AC02-06CH11357.  C.-Y.~C. is
supported by the NSF grant NSF-1740142.  F.~P. and D.~W. are supported
by the DOE grants DE-FG02-91ER40684 and DE-AC02-06CH11357.  This research used resources of the Argonne
Leadership Computing Facility, which is a DOE Office of Science User
Facility supported under Contract DE-AC02-06CH11357.


\appendix
\section{NLO expressions for the helicity fractions}
\label{sc:appendix1}
We present here the analytic results for the NLO SMEFT corrected helicity fractions $F_i$ and the total decay width $\Gamma_\textrm{tot}$ of $t \rightarrow W b$. For convenience we split up the different contributions.
\begin{align}
\Gamma_\textrm{tot} =& \Gamma_\textrm{Born} + \Delta\Gamma_\textrm{QCD}+\Delta\Gamma_\textrm{4f},\nonumber\\
F_i =& F_i^\textrm{Born} + \Delta F_i^\textrm{QCD} + \Delta F_i^\textrm{4f},
\end{align}
where the first term in the expansion describes the contributions coming from the tree-level diagrams only (with full $x_b$ dependence retained), while the second one contains all QCD-like corrections expanded up to order $g_s^2$ and the last the pieces from the four-Fermi operators. We consistently expanded all three contributions to leading order in $\frac{1}{\Lambda^2}$.\\

\subsection{QCD corrected decay fractions}
\begin{align}
\Delta\Gamma_\textrm{QCD}=&-\frac{g_s \overline{g} m_t}{2304 \pi^3 x_W^2}\Big[(x_W^2-1) (6 \sqrt{2} C_{tg} \overline{g}  m_t^2 x_v (1 + 11 x_W^2 - 20 x_W^4)\nonumber\\
& + 
    12 \sqrt{2}
      C_{tW} g_s  m_t^2 x_v x_W^2 (17 - 21 x_W^2 + 4 \pi^2 (x_W^2-1)) +
     g_s \overline{g} (3 (5 + 9 x_W^2 - 6 x_W^4) \nonumber\\
     &+ \pi^2 (-4 - 4 x_W^2 + 
          8 x_W^4))) + 
 \log{(1 - x_W)} (6 (-1 + 
       x_W^2)^2 (4 \sqrt{2} C_{tg} \overline{g}  m_t^2 x_v (x_W^2-1) \nonumber\\
       &+ 
       g_s \overline{g} (5 + 4 x_W^2) + 
       4 \sqrt{2} C_{tW} g_s  m_t^2 x_v (2 + 7 x_W^2)) + 
    24 g_s (x_W^2-1)^2 (\overline{g} + 2 \overline{g} x_W^2\nonumber\\
    & + 
       12 \sqrt{2} C_{tW}  m_t^2 x_v x_W^2) \log{(x_W)}) + 
 6 ( x_W^2-1)^2 (4 \sqrt{2} C_{tg} \overline{g}  m_t^2 x_v (x_W^2-1) \nonumber\\
 &+ 
    g_s \overline{g} (5 + 4 x_W^2) + 
    4 \sqrt{2} C_{tW} g_s  m_t^2 x_v (2 + 7 x_W^2))\log{(1 + x_W)} \nonumber\\
    &+ 
 \log{(x_W)} (24 x_W^2 (4 \sqrt{2} C_{tW} g_s  m_t^2 x_v x_W^2 (3 - 2 x_W^2) + 
       \sqrt{2} C_{tg} \overline{g}  m_t^2 x_v x_W^2 (3 + x_W^2) \nonumber\\
       &- 
       g_s \overline{g} (-1 + x_W^2 + 2 x_W^4)) + 
    24 g_s (x_W^2-1)^2 (\overline{g} + 2 \overline{g} x_W^2 + 
       12 \sqrt{2} C_{tW}  m_t^2 x_v x_W^2) \log{(1 + x_W)}) \nonumber\\
       &+ 
 48 g_s (x_W^2-1)^2 (\overline{g} + 2 \overline{g} x_W^2 + 
    12 \sqrt{2} C_{tW}  m_t^2 x_v x_W^2) \textrm{Li}_2{(-x_W)} \nonumber\\
    &+ 
 48 g_s (-1 + x_W^2)^2 (\overline{g} + 2 \overline{g} x_W^2 + 
    12 \sqrt{2} C_{tW}  m_t^2 x_v x_W^2) \textrm{Li}_2{(x_W)}\Big]
            \label{app:width}
 \end{align}
  
\begingroup
\allowdisplaybreaks
\begin{align}
\Delta F^\textrm{QCD}_{\textrm{L}} =& \frac{g_s}{9\overline{g} \pi^2 (1-x_W^2)^2(1 + 2 x_W^2)^3 }\Big[x_W^2 (-\sqrt{2}
       C_{tg} \overline{g}  m_t^2 x_v (1 + 
       2 x_W^2) (-6 x_W^2 (9 - 10 x_W^2 + x_W^4)\nonumber\\
       & + \pi^2 (1 + 5 x_W^2 + 
          6 x_W^4)) + 
    g_s (\overline{g} (1 + 2 x_W^2) (-\pi^2 (7 + 15 x_W^2 + 2 x_W^4) + 
          6 (6 + 6 x_W^2 \nonumber\\
          & - 13 x_W^4 + x_W^6)) + 
       2 \sqrt{2}
         C_{tW}  m_t^2 x_v (\pi^2 (1 + 49 x_W^2 + 106 x_W^4 + 
             24 x_W^6) + 
          6 (-2 \nonumber\\
          &- 39 x_W^2 - 40 x_W^4 + 79 x_W^6 + 2 x_W^8)))) + 
 3 (1 - x_W)^2  ( (\sqrt{2} C_{tg} \overline{g}  m_t^2 x_v (4 x_W^4-1) \nonumber\\
 &+ 
       g_s (\overline{g} + 2 \overline{g} x_W^2 + 
          2 \sqrt{2} C_{tW}  m_t^2 x_v (1 - 10 x_W^2)))(1 + x_W)^2 (1 - 
       x_W^2) \nonumber\\
       &- (1 - 
       x_W) x_W (1 + 
       2 x_W^2) (4 \sqrt{2} C_{tW} g_s  m_t^2 x_v x_W (-1 + 15 x_W + x_W^2)\nonumber\\
       & + 
       2 \sqrt{2} C_{tg} \overline{g}  m_t^2 x_v x_W (1 + 2 x_W^2) - 
       g_s \overline{g} (5 + x_W + 10 x_W^2 + 2 x_W^3)) \log{(x_W)})\log{(1 - x_W)}\nonumber\\
       & + 
 3  (1 - x_W^2)^3 (\sqrt{2} C_{tg} \overline{g}  m_t^2 x_v (4 x_W^4-1) + 
    g_s (\overline{g} + 2 \overline{g} x_W^2 + 
       2 \sqrt{2} C_{tW}  m_t^2 x_v (1 \nonumber\\
       &- 10 x_W^2))) \log{(1 + x_W)} + 
 3 x_W \log{(x_W)} (2 x_W (\sqrt{2}
         C_{tg} \overline{g}  m_t^2 x_v x_W^2 (7 + 23 x_W^2 + 18 x_W^4)\nonumber\\
         & - 
       2 \sqrt{2}
         C_{tW} g_s  m_t^2 x_v x_W^2 (35 + 101 x_W^2 + 112 x_W^4 + 4 x_W^6) +
        g_s \overline{g} (5 + 25 x_W^2+ 44 x_W^4  \nonumber\\
        &+ 28 x_W^6)) - (1 + x_W)^3 (1 + 
       2 x_W^2) (4 \sqrt{2} C_{tW} g_s  m_t^2 x_v x_W (-1 - 15 x_W + x_W^2)\nonumber\\
       & + 
       2 \sqrt{2} C_{tg} \overline{g}  m_t^2 x_v x_W (1 + 2 x_W^2) - 
       g_s \overline{g} (-5 + x_W - 10 x_W^2 + 2 x_W^3)) \log{(1 + x_W)})\nonumber\\
       & - 
 3 x_W (1 + x_W)^3 (1 + 
    2 x_W^2) (4 \sqrt{2} C_{tW} g_s  m_t^2 x_v x_W (-1 - 15 x_W + x_W^2)\nonumber\\
    & + 
    2 \sqrt{2} C_{tg} \overline{g}  m_t^2 x_v x_W (1 + 2 x_W^2) + 
    g_s \overline{g} (5 - x_W + 10 x_W^2 - 2 x_W^3)) \textrm{Li}_2{(-x_W)} \nonumber\\
    &- 
 3 (1 - x_W)^3 x_W (1 + 
    2 x_W^2) (4 \sqrt{2} C_{tW} g_s  m_t^2 x_v x_W (-1 + 15 x_W + x_W^2)\nonumber\\
    & + 
    2 \sqrt{2} C_{tg} \overline{g}  m_t^2 x_v x_W (1 + 2 x_W^2) - 
    g_s \overline{g} (5 + x_W + 10 x_W^2 + 2 x_W^3)) \textrm{Li}_2{(x_W)}\Big]
\label{app:fl}
\end{align}

\begin{align}
\Delta F^\textrm{QCD}_{-} =& \frac{g_s}{36 \overline{g} \pi^2 (1 - x_W^2)^2 (1 + 2 x_W^2)^3}\Big[x_W^2 (6 \sqrt{2}
      \overline{g}  m_t^2 x_v (1 - x_W + 2 x_W^2 - 
       2 x_W^3) ( 
       2 C_{tg} (2 - 3 x_W\nonumber\\
       & - 6 x_W^2 - 23 x_W^3 + 18 x_W^4 + 4 x_W^5)+C_{bg} (1 + x_W)^2 (-1 + x_W - 2 x_W^2 + 2 x_W^3) ) \nonumber\\
       &+ 
    g_s (\overline{g} (1 + 2 x_W^2) (2 \pi^2 (7 + 10 x_W^2 - 6 x_W^4 + 4 x_W^6) - 
          3 (23 + 20 x_W - 6 x_W^2 + 48 x_W^3 \nonumber\\
          &- 111 x_W^4 + 16 x_W^5 + 
             10 x_W^6)) - 
       4 \sqrt{2}
         C_{tW}  m_t^2 x_v (3 (-8 + 30 x_W - 133 x_W^2 + 82 x_W^3\nonumber\\
         & - 
             206 x_W^4 + 64 x_W^5 + 129 x_W^6 + 40 x_W^7 + 2 x_W^8) + 
          2 \pi^2 (1 + 33 x_W^2 + 64 x_W^4 + 6 x_W^6 + 4 x_W^8))))\nonumber\\ 
          &+ 
 \log{(1 - x_W)} (12 (x_W^2 - 
       1)^3 (\sqrt{2} C_{tg} \overline{g}  m_t^2 x_v (-1 + 4 x_W^4) + 
       g_s (\overline{g} + 2 \overline{g} x_W^2\nonumber\\
       & + 
          2 \sqrt{2} C_{tW}  m_t^2 x_v (1 - 10 x_W^2))) + 
    6 (x_W-1)^2 x_W (1 + 
       2 x_W^2) (2 \sqrt{2}
         C_{tg} \overline{g}  m_t^2 x_v x_W (1\nonumber\\
         & - x_W + 2 x_W^2 - 2 x_W^3) - 
       4 \sqrt{2}
         C_{tW} g_s  m_t^2 x_v x_W (-3 - 24 x_W + 14 x_W^2 + 9 x_W^3 + 
          4 x_W^4) \nonumber\\
          &+ 
       g_s \overline{g} (-5 + 8 x_W - x_W^2 + 20 x_W^3 + 18 x_W^4 + 8 x_W^5)) \log{(
      x_W)}) \nonumber\\
      &+ 12 (x_W^2 - 
    1) (-\sqrt{2} C_{tg} \overline{g}  m_t^2 x_v (1 + x_W^2 + 10 x_W^4 + 24 x_W^6) + 
    g_s \overline{g} (1 - 5 x_W^2- 21 x_W^4 \nonumber\\
    & - 10 x_W^6 + 8 x_W^8) + 
    2 \sqrt{2}
      C_{tW} g_s  m_t^2 x_v (1 - 25 x_W^2 + 2 x_W^4 - 2 x_W^6 - 
       12 x_W^8)) \log{(1 + x_W)}\nonumber\\
       & + 
 \log{(x_W)} (-12 (\sqrt{2}
         C_{tg} \overline{g}  m_t^2 x_v x_W^4 (7 + 17 x_W^2 + 4 x_W^4 - 4 x_W^6)\nonumber\\
         & - 
       2 \sqrt{2}
         C_{tW} g_s  m_t^2 x_v x_W^4 (39 + 93 x_W^2 + 136 x_W^4 + 20 x_W^6)+
        g_s \overline{g} x_W^2 (5 + 23 x_W^2 + 42 x_W^4 \nonumber\\
        &+ 36 x_W^6 + 8 x_W^8)) + 
    6 x_W (1 + x_W)^2 (1 + 
       2 x_W^2) (2 \sqrt{2}
         C_{tg} \overline{g}  m_t^2 x_v x_W (1 + x_W \nonumber\\
         &+ 2 x_W^2 + 2 x_W^3) - 
       4 \sqrt{2}
         C_{tW} g_s  m_t^2 x_v x_W (-3 + 24 x_W + 14 x_W^2 - 9 x_W^3 + 
          4 x_W^4) \nonumber\\
          &+ 
       g_s \overline{g} (5 + 8 x_W + x_W^2 + 20 x_W^3 - 18 x_W^4 + 8 x_W^5))\log{(
      1 + x_W)})\nonumber\\
      & + 
 6 x_W (1 + 
    2 x_W^2) (2 \sqrt{2}
      C_{tg} \overline{g}  m_t^2 x_v (-1 + x_W)^3 x_W (1 + 2 x_W^2) + 
    4 \sqrt{2}
      C_{tW} g_s  m_t^2 x_v x_W (1\nonumber\\
      & - 18 x_W - 89 x_W^2 - 43 x_W^3 + 16 x_W^4 +
        x_W^5 - 12 x_W^6) + 
    g_s \overline{g} (5 + 18 x_W + 22 x_W^2 \nonumber\\
    &+ 10 x_W^3 + 23 x_W^4 - 40 x_W^5 - 
       2 x_W^6 + 24 x_W^7)) \textrm{Li}_2{(-x_W)} \nonumber\\
       &+ 
 6 (1 - x_W)^2 x_W (1 + 
    2 x_W^2) (-2 \sqrt{2}
      C_{tg} \overline{g}  m_t^2 x_v x_W (-1 + x_W - 2 x_W^2 + 2 x_W^3)\nonumber\\
      & - 
    4 \sqrt{2}
      C_{tW} g_s  m_t^2 x_v x_W (-3 - 24 x_W + 14 x_W^2 + 9 x_W^3 + 4 x_W^4) +
     g_s \overline{g} (-5 + 8 x_W\nonumber\\
     & - x_W^2 + 20 x_W^3 + 18 x_W^4 + 8 x_W^5)) \textrm{Li}_2{(x_W)}\Big]
   \label{app:fm}
\end{align}
\endgroup

\subsection{Four-fermion corrected decay fractions}
Here we report the four-Fermi corrected longitudinal and negative transverse helicity fractions, and the total decay width of $t \rightarrow W b$. We omit the contributions stemming from the $t-b-u-d$ vertex leading to two massless quarks in the loop.

\begin{align}
\Delta\Gamma_\textrm{4f} =&\frac{\overline{g}m_t^3(x_W^2-1)^2}{4608\pi^2x_W^6}\Big[12 C_{lq}^{(3)} x_W^6 (1 + 2 x_W^2)\log{(x_W^2)} - 
 x_W^2 (x_W^4 (20 C_{lq}^{(3)} + 27 C_{qu}^{(1)} + 36 C_{qu}^{(8)} \nonumber\\
 &+ 40 C_{lq}^{(3)} x_W^2) + 
    2 C_{qq}^{(1)} (-3 - 3 x_W^2 + 10 x_W^4 + 8 x_W^6) + 
    2 C_{qq}^{(3)} (-15 - 69 x_W^2 - 22 x_W^4\nonumber\\
  & + 112 x_W^6)) + 
 6 (C_{qq}^{(1)} + 5 C_{qq}^{(3)}) (1 + x_W^2 - 2 x_W^4)^2 \log{(1 - x_W^2)}\Big]
 \end{align}

\begin{align}
\Delta F^\textrm{4f}_{\textrm{L}} =& \frac{m_t^2 x_W^2}{12 \pi^2}\frac{(x_W^2-1)}{(1 + 2 x_W^2)^2}(3 C_{qu}^{(1)} + 4 C_{qu}^{(8)})
\end{align}

\begin{align}
\Delta F^\textrm{4f}_{\textrm{-}} = &- \frac{m_t^2 x_W^2}{12 \pi^2}\frac{(x_W^2-1)}{(1 + 2 x_W^2)^2}(3 C_{qu}^{(1)} + 4 C_{qu}^{(8)})
\end{align}



\begin{thebibliography}{99}
\frenchspacing

\bibitem{Buchmuller:1985jz} 
  W.~Buchmuller and D.~Wyler,
  Nucl.\ Phys.\ B {\bf 268}, 621 (1986).
  doi:10.1016/0550-3213(86)90262-2

\bibitem{Grzadkowski:2010es} 
  B.~Grzadkowski, M.~Iskrzynski, M.~Misiak and J.~Rosiek,
  JHEP {\bf 1010}, 085 (2010)
  doi:10.1007/JHEP10(2010)085
  [arXiv:1008.4884 [hep-ph]].

\bibitem{Han:2004az} 
  Z.~Han and W.~Skiba,
  Phys.\ Rev.\ D {\bf 71}, 075009 (2005)
  doi:10.1103/PhysRevD.71.075009
  [hep-ph/0412166].

\bibitem{Pomarol:2013zra} 
  A.~Pomarol and F.~Riva,
  JHEP {\bf 1401}, 151 (2014)
  doi:10.1007/JHEP01(2014)151
  [arXiv:1308.2803 [hep-ph]].
  
\bibitem{Chen:2013kfa} 
  C.~Y.~Chen, S.~Dawson and C.~Zhang,
  Phys.\ Rev.\ D {\bf 89}, no. 1, 015016 (2014)
  doi:10.1103/PhysRevD.89.015016
  [arXiv:1311.3107 [hep-ph]].

\bibitem{Ellis:2014dva} 
  J.~Ellis, V.~Sanz and T.~You,
  JHEP {\bf 1407}, 036 (2014)
  doi:10.1007/JHEP07(2014)036
  [arXiv:1404.3667 [hep-ph]].
  
\bibitem{Wells:2014pga} 
  J.~D.~Wells and Z.~Zhang,
  Phys.\ Rev.\ D {\bf 90}, no. 3, 033006 (2014)
  doi:10.1103/PhysRevD.90.033006
  [arXiv:1406.6070 [hep-ph]].

 \bibitem{Falkowski:2014tna} 
  A.~Falkowski and F.~Riva,
  JHEP {\bf 1502}, 039 (2015)
  doi:10.1007/JHEP02(2015)039
  [arXiv:1411.0669 [hep-ph]].
  
\bibitem{deBlas:2016ojx} 
  J.~de Blas, M.~Ciuchini, E.~Franco, S.~Mishima, M.~Pierini, L.~Reina and L.~Silvestrini,
  JHEP {\bf 1612}, 135 (2016)
  doi:10.1007/JHEP12(2016)135
  [arXiv:1608.01509 [hep-ph]].

\bibitem{Hartmann:2015oia} 
  C.~Hartmann and M.~Trott,
  JHEP {\bf 1507}, 151 (2015)
  doi:10.1007/JHEP07(2015)151
  [arXiv:1505.02646 [hep-ph]].

 \bibitem{Hartmann:2015aia} 
  C.~Hartmann and M.~Trott,
  Phys.\ Rev.\ Lett.\  {\bf 115}, no. 19, 191801 (2015)
  doi:10.1103/PhysRevLett.115.191801
  [arXiv:1507.03568 [hep-ph]].

\bibitem{Dedes:2018seb} 
  A.~Dedes, M.~Paraskevas, J.~Rosiek, K.~Suxho and L.~Trifyllis,
  JHEP {\bf 1808}, 103 (2018)
  doi:10.1007/JHEP08(2018)103
  [arXiv:1805.00302 [hep-ph]].

\bibitem{Gauld:2015lmb} 
  R.~Gauld, B.~D.~Pecjak and D.~J.~Scott,
  JHEP {\bf 1605}, 080 (2016)
  doi:10.1007/JHEP05(2016)080
  [arXiv:1512.02508 [hep-ph]].

\bibitem{Gauld:2016kuu} 
  R.~Gauld, B.~D.~Pecjak and D.~J.~Scott,
  Phys.\ Rev.\ D {\bf 94}, no. 7, 074045 (2016)
  doi:10.1103/PhysRevD.94.074045
  [arXiv:1607.06354 [hep-ph]].

 \bibitem{Dawson:2018pyl} 
  S.~Dawson and P.~P.~Giardino,
  Phys.\ Rev.\ D {\bf 97}, no. 9, 093003 (2018)
  doi:10.1103/PhysRevD.97.093003
  [arXiv:1801.01136 [hep-ph]].

 \bibitem{Dawson:2018liq} 
  S.~Dawson and P.~P.~Giardino,
  Phys.\ Rev.\ D {\bf 98}, no. 9, 095005 (2018)
  doi:10.1103/PhysRevD.98.095005
  [arXiv:1807.11504 [hep-ph]].

\bibitem{Hartmann:2016pil} 
  C.~Hartmann, W.~Shepherd and M.~Trott,
  JHEP {\bf 1703}, 060 (2017)
  doi:10.1007/JHEP03(2017)060
  [arXiv:1611.09879 [hep-ph]].
  
\bibitem{Dawson:2018jlg} 
  S.~Dawson and A.~Ismail,
  Phys.\ Rev.\ D {\bf 98}, no. 9, 093003 (2018)
  doi:10.1103/PhysRevD.98.093003
  [arXiv:1808.05948 [hep-ph]].
  
\bibitem{Dawson:2018dxp} 
  S.~Dawson, P.~P.~Giardino and A.~Ismail,
  arXiv:1811.12260 [hep-ph].

 \bibitem{Degrande:2012gr} 
  C.~Degrande, J.~M.~Gerard, C.~Grojean, F.~Maltoni and G.~Servant,
  JHEP {\bf 1207}, 036 (2012)
  Erratum: [JHEP {\bf 1303}, 032 (2013)]
  doi:10.1007/JHEP07(2012)036, 10.1007/JHEP03(2013)032
  [arXiv:1205.1065 [hep-ph]].
  
\bibitem{Vryonidou:2018eyv} 
  E.~Vryonidou and C.~Zhang,
  JHEP {\bf 1808}, 036 (2018)
  doi:10.1007/JHEP08(2018)036
  [arXiv:1804.09766 [hep-ph]].

\bibitem{Berthier:2015oma} 
  L.~Berthier and M.~Trott,
  JHEP {\bf 1505}, 024 (2015)
  doi:10.1007/JHEP05(2015)024
  [arXiv:1502.02570 [hep-ph]].

\bibitem{Passarino:2016pzb} 
  G.~Passarino and M.~Trott,
  arXiv:1610.08356 [hep-ph].

\bibitem{Greiner:2011tt} 
  N.~Greiner, S.~Willenbrock and C.~Zhang,
  Phys.\ Lett.\ B {\bf 704}, 218 (2011)
  doi:10.1016/j.physletb.2011.09.026
  [arXiv:1104.3122 [hep-ph]].

\bibitem{Zhang:2012cd} 
  C.~Zhang, N.~Greiner and S.~Willenbrock,
  Phys.\ Rev.\ D {\bf 86}, 014024 (2012)
  doi:10.1103/PhysRevD.86.014024
  [arXiv:1201.6670 [hep-ph]].

  
\bibitem{Buckley:2015lku} 
  A.~Buckley, C.~Englert, J.~Ferrando, D.~J.~Miller, L.~Moore, M.~Russell and C.~D.~White,
  JHEP {\bf 1604}, 015 (2016)
  doi:10.1007/JHEP04(2016)015
  [arXiv:1512.03360 [hep-ph]].

\bibitem{Cirigliano:2016nyn} 
  V.~Cirigliano, W.~Dekens, J.~de Vries and E.~Mereghetti,
  Phys.\ Rev.\ D {\bf 94}, no. 3, 034031 (2016)
  doi:10.1103/PhysRevD.94.034031
  [arXiv:1605.04311 [hep-ph]].
  
\bibitem{AguilarSaavedra:2018nen} 
  J.~A.~Aguilar-Saavedra {\it et al.},
  arXiv:1802.07237 [hep-ph].
  
\bibitem{Hartland:2019bjb} 
  N.~P.~Hartland, F.~Maltoni, E.~R.~Nocera, J.~Rojo, E.~Slade, E.~Vryonidou and C.~Zhang,
  arXiv:1901.05965 [hep-ph].

\bibitem{Zhang:2014rja} 
  C.~Zhang,
  Phys.\ Rev.\ D {\bf 90}, no. 1, 014008 (2014)
  doi:10.1103/PhysRevD.90.014008
  [arXiv:1404.1264 [hep-ph]].

\bibitem{Azzi:2019yne} 
  P.~Azzi {\it et al.} [HL-LHC Collaboration and HE-LHC Working Group],
  arXiv:1902.04070 [hep-ph].

\bibitem{Dedes:2017zog} 
  A.~Dedes, W.~Materkowska, M.~Paraskevas, J.~Rosiek and K.~Suxho,
  JHEP {\bf 1706}, 143 (2017)
  doi:10.1007/JHEP06(2017)143
  [arXiv:1704.03888 [hep-ph]].
  
  \bibitem{Jenkins:2013zja} 
  E.~E.~Jenkins, A.~V.~Manohar and M.~Trott,
  JHEP {\bf 1310}, 087 (2013)
  doi:10.1007/JHEP10(2013)087
  [arXiv:1308.2627 [hep-ph]].
  
  \bibitem{Jenkins:2013wua} 
  E.~E.~Jenkins, A.~V.~Manohar and M.~Trott,
  JHEP {\bf 1401}, 035 (2014)
  doi:10.1007/JHEP01(2014)035
  [arXiv:1310.4838 [hep-ph]].

 \bibitem{Alonso:2013hga} 
  R.~Alonso, E.~E.~Jenkins, A.~V.~Manohar and M.~Trott,
  JHEP {\bf 1404}, 159 (2014)
  doi:10.1007/JHEP04(2014)159
  [arXiv:1312.2014 [hep-ph]].

\bibitem{Fischer:2000kx} 
  M.~Fischer, S.~Groote, J.~G.~Korner and M.~C.~Mauser,
  Phys.\ Rev.\ D {\bf 63}, 031501 (2001)
  doi:10.1103/PhysRevD.63.031501
  [hep-ph/0011075].

 \bibitem{Do:2002ky} 
  H.~S.~Do, S.~Groote, J.~G.~Korner and M.~C.~Mauser,
  Phys.\ Rev.\ D {\bf 67}, 091501 (2003)
  doi:10.1103/PhysRevD.67.091501
  [hep-ph/0209185].
  
  
\bibitem{Chetyrkin:1981qh} 
  K.~G.~Chetyrkin and F.~V.~Tkachov,
  Nucl.\ Phys.\ B {\bf 192}, 159 (1981).
  doi:10.1016/0550-3213(81)90199-1

\bibitem{Jegerlehner:2000dz} 
  F.~Jegerlehner,
  Eur.\ Phys.\ J.\ C {\bf 18}, 673 (2001)
  doi:10.1007/s100520100573
  [hep-th/0005255].

  \bibitem{Trueman:1995ca} 
  T.~L.~Trueman,
  Z.\ Phys.\ C {\bf 69}, 525 (1996)
  doi:10.1007/BF02907437
  [hep-ph/9504315].

\bibitem{Larin:1993tq} 
  S.~A.~Larin,
  Phys.\ Lett.\ B {\bf 303}, 113 (1993)
  doi:10.1016/0370-2693(93)90053-K
  [hep-ph/9302240].

  \bibitem{Breitenlohner:1975hg} 
  P.~Breitenlohner and D.~Maison,
  Commun.\ Math.\ Phys.\  {\bf 52}, 39 (1977).
  doi:10.1007/BF01609070

\bibitem{Broadhurst:1994se} 
  D.~J.~Broadhurst and A.~G.~Grozin,
  Phys.\ Rev.\ D {\bf 52}, 4082 (1995)
  doi:10.1103/PhysRevD.52.4082
  [hep-ph/9410240].

  \bibitem{Donoghue:1992dd} 
  J.~F.~Donoghue, E.~Golowich and B.~R.~Holstein,
  Camb.\ Monogr.\ Part.\ Phys.\ Nucl.\ Phys.\ Cosmol.\  {\bf 2}, 1 (1992)
  [Camb.\ Monogr.\ Part.\ Phys.\ Nucl.\ Phys.\ Cosmol.\  {\bf 35} (2014)], Section III-3.
  
 \bibitem{Shtabovenko:2016olh} 
  V.~Shtabovenko,
  J.\ Phys.\ Conf.\ Ser.\  {\bf 762}, no. 1, 012064 (2016)
  doi:10.1088/1742-6596/762/1/012064
  [arXiv:1604.06709 [hep-ph]].
  
  \bibitem{Jamin:1991dp} 
  M.~Jamin and M.~E.~Lautenbacher,
  Comput.\ Phys.\ Commun.\  {\bf 74}, 265 (1993).
  doi:10.1016/0010-4655(93)90097-V
  
    
  \bibitem{Gamma5tests} 
  A.~V.~Bednyakov, A.~F.~Pikelner and V.~N.~Velizhanin,
  J.\ Phys.\ Conf.\ Ser.\  {\bf 523}, 012045 (2014)
  doi:10.1088/1742-6596/523/1/012045
  [arXiv:1309.1643 [hep-ph]];\\
   A.~V.~Bednyakov, A.~F.~Pikelner and V.~N.~Velizhanin,
  Phys.\ Lett.\ B {\bf 722}, 336 (2013)
  doi:10.1016/j.physletb.2013.04.038
  [arXiv:1212.6829 [hep-ph]];\\
   D.~St\"ockinger, Dimensional regularization and $\gamma_5$, FCCee workshop talk (2018)"
  

  \bibitem{Adel:1994my} 
  K.~Adel and Y.~P.~Yao,
  Phys.\ Rev.\ D {\bf 53}, 374 (1996)
  doi:10.1103/PhysRevD.53.374
  [hep-ph/9408341].
  
  \bibitem{Herrlich:1994kh} 
  S.~Herrlich and U.~Nierste,
  Nucl.\ Phys.\ B {\bf 455}, 39 (1995)
  doi:10.1016/0550-3213(95)00474-7
  [hep-ph/9412375].

   \bibitem{Caswell:1974gg} 
  W.~E.~Caswell,
  Phys.\ Rev.\ Lett.\  {\bf 33}, 244 (1974).

\bibitem{Tanabashi:2018oca} 
  M.~Tanabashi {\it et al.} [Particle Data Group],
  Phys.\ Rev.\ D {\bf 98}, no. 3, 030001 (2018).
  doi:10.1103/PhysRevD.98.030001

\bibitem{Khachatryan:2016fky} 
  V.~Khachatryan {\it et al.} [CMS Collaboration],
  Phys.\ Lett.\ B {\bf 762}, 512 (2016)
  doi:10.1016/j.physletb.2016.10.007
  [arXiv:1605.09047 [hep-ex]].

   \bibitem{atlas-note}
   ATLAS Collaboration, ATL-PHYS-PUB-2019-005.

   \bibitem{cms-note}
   CMS Collaboration, CMS NOTE 2018-006.
   
\bibitem{Zhang:2010dr} 
  C.~Zhang and S.~Willenbrock,
  Phys.\ Rev.\ D {\bf 83}, 034006 (2011)
  doi:10.1103/PhysRevD.83.034006
  [arXiv:1008.3869 [hep-ph]].

 
  \bibitem{Benedikt:2018qee} 
  A.~Abada {\it et al.} [FCC Collaboration],
  CERN-ACC-2018-0057.

\end{thebibliography}
\end{document}